\documentclass[11pt]{article}
\usepackage[T2A]{fontenc}       
\usepackage[utf8]{inputenc}     
\usepackage[russian,english]{babel}
\usepackage{amssymb,amsfonts,amsmath,mathtext}
\usepackage{cite,enumerate,float,indentfirst}
\usepackage[vcentermath]{youngtab}
\usepackage{tikz}
\usepackage[centertableaux,smalltableaux]{ytableau}
\usepackage{theorem,cite,epsfig,url,psfrag,eepic,amsmath}
\usepackage[colorlinks,allcolors=blue]{hyperref}
\usepackage{wrapfig}
\usepackage{empheq}
\usepackage{graphicx}
\usepackage{pstricks}
\usepackage[symbol]{footmisc}

\let\oldsection\section
\renewcommand{\section}{
\renewcommand{\theequation}{\thesection.\arabic{equation}}
\oldsection}

\newcommand{\beq}{\begin{equation}}
\newcommand{\eeq}{\end{equation}}
\newcommand{\beqq}{\begin{equation*}}
\newcommand{\eeqq}{\end{equation*}}
\newcommand\beqa{\begin{eqnarray}}
\newcommand\eeqa{\end{eqnarray}}
\newcommand\bea{\begin{array}}
\newcommand\eea{\end{array}}

\def\XXint#1#2#3{{\setbox0=\hbox{$#1{#2#3}{\int}$}
\vcenter{\hbox{$#2#3$}}\kern-.5\wd0}}

\def\[{\left[}
\def\]{\right]}

\def\spi{\relax{\rm \pi\kern-0.5em /}}
\def\sA{\relax{\rm A\kern-0.5em /}}
\def\sp{\relax{\rm p\kern-0.5em /}}
\def\sd{\relax{\rm \d\kern-0.5em /}}
\def\sk{\relax{\rm k\kern-0.5em /}}
\def\sn{\relax{\rm n\kern-0.5em /}}
\def\sl{\relax{\rm l\kern-0.5em /}}
\def\sP{\relax{\rm P\kern-0.7em /}}
\def\sBethe{\relax{\rm \Bethe\kern-0.5em /}}

\usepackage{varioref}
\usepackage{makeidx}
\makeindex

\newcommand\blfootnote[1]{%
  \begingroup
  \renewcommand\thefootnote{}\footnote{\hspace{-6mm}#1}%
  \addtocounter{footnote}{-1}%
  \endgroup
}

\par\vspace{1.5cm}
\vspace{2mm}

\begin{document}
\flushbottom

\begin{center}
    
{\Large\textbf{\mathversion{bold}
On Integrable Deformations of CFTs and Integrable Two-Dimensional Sigma Models
}\par}
\vspace{1.0cm}

\textrm{M.N. Alfimov$^\text{\tiny a,b}$\footnote[2]{\href{mailto:malfimov@hseru}{M. Alfimov}}}
\blfootnote{malfimov AT hse.ru}
\\ \vspace{1.2cm}
\footnotesize{
\textit{${}^\text{\tiny a}$ National Research University Higher School Economics, \\
6 Usacheva St., Moscow 119048, Russia}
\\
\textit{${}^\text{\tiny b}$ P.N. Lebedev Physical Institute of the Russian Academy of Sciences, \\
53 Leninsky Prospekt, Moscow 119991, Russia}
\\
\vspace{4mm}
}
\end{center}

\begin{abstract}
In this paper, we analyze the systems of screening charges corresponding to $\eta$-deformed $O(N)$ and $OSp(N|2m)$ sigma models. These sigma models can be represented as integrable deformations of certain conformal field theories. Based on the case of $O(N)$ sigma models, which were conjectured in previous works by A.V. Litvinov and co-authors to be an integrable deformation of the conformal field theory corresponding to the algebras $W(D(n+1|n))$ and $W(D(n|n))$ for the cases $N=2n+2$ and $N=2n+1$ at $n \geq 1$, respectively, we formulate a hypothesis on how to write down the system of screening charges for the conformal field theory whose integrable deformation defines the $\eta$-deformed $OSp(N|2m)$ sigma model for the case of odd $N=2n+1$, asymptotically free models with $N-2m-2>0$ and $n \geq 2$, $m \geq 1$. This hypothesis is explicitly verified for the cases of $OSp(5|2)$, $OSp(7|2)$, and $OSp(7|4)$, which were separately investigated in the work of M.N. Alfimov and co-authors.
\end{abstract}

\section{Introduction}

It is well known that integrability-preserving deformations of $O(N)$ sigma models possess a dual description in terms of an associated theory of bosons and fermions with exponential interactions of the Toda type \cite{Fateev:1992tk,Fateev:1996ea,Litvinov:2018bou}. The essence of this approach relies on the fact that if we take the trigonometric $O(N)$ $S$-matrix \cite{Bazhanov:1984gu,Jimbo:1985ua}, which is a solution to the Yang-Baxter equation, and consider its value at the deformation parameter value $q=-1$, it turns out to be proportional to the graded permutation operator. Physically, this means that the theory becomes free at this point and consists of a set of free bosonic and fermionic fields. Also, the renormalization group flow of trigonometrically deformed $O(N)$ sigma models is structured such that in the ultraviolet limit, the geometry of the field space becomes cylindrical; that is, it is apparent that the sigma model flows to a free theory at this point. If one looks at the corrections to this free limit for the sigma model action, it turns out that they represent relevant exponential operators or Wakimoto-type screenings \cite{Wakimoto:1986gf}, which allows one to conjecture that the sigma model can be viewed as an integrable perturbation of a certain free field theory by such operators. As shown in \cite{Litvinov:2016mgi}, the set of Wakimoto-type screenings can be replaced by a set of screenings that determines the same quantum integrals of motion and, thus, provides a dual description of the theory under consideration. From this description in terms of a Toda-type theory, one can reconstruct the aforementioned $S$-matrix order by order in the deviation from the critical value of the deformation parameter $q=-1$.

On the other hand, there are known examples of integrable superstring theories, such as the $\textrm{AdS}_5 \times \textrm{S}^5$ type IIB superstring (dual to $\mathcal{N}=4$ SYM) and others, which also possess integrable deformations. At the same time, the corresponding $q$-deformed $S$-matrices, for instance \cite{Hoare:2011wr}, presumably enjoy the same property, i.e., they are proportional to the graded permutation operator at the point $q=-1$. Our strategic goal, outlined in \cite{Alfimov:2020jpy}, is to construct a similar dual description for the deformed $\textrm{AdS}_5 \times \textrm{S}^5$ type IIB superstring \cite{Arutyunov:2013ega,Arutyunov:2015qva,Arutyunov:2015mqj} and possibly other theories of this type. However, there are at least three significant obstacles along this path compared to $O(N)$ and similar bosonic models:
\begin{enumerate}

\item Incorporating the fermionic degrees of freedom present in superstring theory into the construction of the dual Toda-type theory.

\item Adapting the entire mechanism of the dual description for application to sigma models with a non-compact target space.

\item Superstring theory possesses reparametrization invariance and requires gauge fixing, which implies incorporating this symmetry into the dual description.

\end{enumerate}
In this article, we intend to focus on the general scheme for constructing the dual description of deformed $O(N)$ and $OSp(N|2m)$ sigma models, where the latter can be viewed as the simplest generalization of sigma models on a sphere that contains fermionic degrees of freedom. In previous works \cite{Litvinov:2018bou,Alfimov:2020jpy}, a number of principles were formulated upon which the dual Toda-type theory is built:

\begin{enumerate}

\item The sigma model under consideration must be renormalizable in the Friedan sense \cite{Friedan:1980jf,Friedan:1980jm}, and its target-space metric must satisfy the Ricci flow equation at the one-loop order. For the models we consider, the Lagrangian in the ultraviolet limit flows either to a free or to a conformal field theory. In the case of deformed $O(N)$ and $OSp(N|2m)$, this can be easily verified by solving the renormalization group flow equation.

\item The dual theory is sought as an integrable deformation from the ``free'' point, where the $S$-matrix solving the Yang-Baxter equations equals the graded permutation operator, and is determined by a set of screening charges that commute with the integrals of motion in the leading order with respect to the mass parameter
\begin{equation}
\[I_k^{\textrm{free}},\int e^{(\boldsymbol{\alpha}_r,\phi)} dz\]=0\;,
\end{equation}

where $\boldsymbol{\alpha}_r$ are the lengths of the vectors defining the screening charges in the space of scalar fields.
\item In the general case, the Toda-type theory specifying the dual description of a deformed sigma model can be represented as a conformal field theory defined by screening charges, perturbed by additional screenings so that the resulting theory is integrable. For instance, in the case of the $\eta$-deformed $O(3)$ model, the screenings look like $e^{b\Phi+i\beta\varphi}$, $e^{b\Phi-i\beta\varphi}$, $e^{-b\Phi+i\beta\varphi}$, and $e^{-b\Phi-i\beta\varphi}$, where $b$ is a certain continuous parameter and $\beta=\sqrt{1+b^2}$. On the other hand, for example, the two operators $e^{b\Phi+i\beta\varphi}$ and $e^{b\Phi-i\beta\varphi}$ define the sine-Liouville conformal field theory; thus, the dual description can be understood as an integrable perturbation of this conformal field theory. It follows that our $O(N)$ and $OSP(N|2m)$ models are integrable deformations of certain conformal field theories defined by the cosets of affine Lie algebras
\beq
\frac{\widehat{\mathfrak{so}}(N)_w}{\widehat{\mathfrak{so}}(N-1)_w} \quad \mathrm{and} \quad \frac{\widehat{\mathfrak{osp}}(N|2m)_w}{\widehat{\mathfrak{osp}}(N-1|2m)_w};,
\eeq
respectively, where $w$ is a certain level depending on the parameters $b$, $N$, and $m$. In what follows, we will mostly omit indicating its explicit value for brevity of notation.
\end{enumerate}
In the next section, we intend to remind the reader how various conformal field theories can be specified using a set of so-called screening charges, and we will also consider transformations of screening charge systems that do not alter the theory they define.

\section{Conformal field theories defined by screening charges}

In this section, we mainly follow the presentation given in works \cite{Litvinov:2016mgi,Litvinov:2018bou}. Let us consider the properties of a conformal field theory defined by a set of exponential screening charges, starting without prefactors. Let $\boldsymbol{\varphi}(z)=(\varphi_{1}(z),\dots,\varphi_{N}(z))$ be an $N$-component holomorphic bosonic field normalized by the operator product expansion
\beq
   \varphi_{i}(z)\varphi_{j}(z')=-\delta_{ij}\log(z-z')+\dots \quad \textrm{as} \quad z \rightarrow z'\;,
\eeq
and $\vec{\boldsymbol{\alpha}}=(\boldsymbol{\alpha}_{1},\dots,\boldsymbol{\alpha}_{N})$ be a set of linearly independent vectors in the field space. We define the $W_{\vec{\boldsymbol{\alpha}}}$-algebra as a set of integer-spin currents $W_{s}(z)$ such that
\beq\label{comm_cond}
  \oint\limits_{\mathcal{C}_{z}}e^{(\boldsymbol{\alpha}_{r}\cdot\boldsymbol{\varphi}(\xi))}W_{s}(z)d\xi=0\;, \quad r=1,\ldots,N\;,
\eeq
where $\mathcal{C}_z$ is a certain contour in the complex plane enclosing the point $z$. The existence of such currents depends on the values of the vectors $\boldsymbol{\alpha}_r$ and the relations between them.

For arbitrary $\vec{\boldsymbol{\alpha}}$, there always exists a spin-$2$ current
\beq
   W_{2}(z)=-\frac{1}{2}(\partial\boldsymbol{\varphi}(z)\cdot\partial\boldsymbol{\varphi}(z))+(\boldsymbol{\rho}\cdot\partial^{2}\boldsymbol{\varphi}(z))\;, \quad \boldsymbol{\rho}=\sum_{r=1}^{N}\left(1+\frac{(\boldsymbol{\alpha}_{r}\cdot\boldsymbol{\alpha}_{r})}{2}\right)\hat{\boldsymbol{\alpha}}_{r}\;,
\eeq
and $(\boldsymbol{\alpha}_{r}\cdot\hat{\boldsymbol{\alpha}}_{s})=\delta_{r,s}$, meaning that $\hat{\boldsymbol{\alpha}}_s$ are the vectors dual to the vectors $\boldsymbol{\alpha}_r$. The corresponding central charge equals
\beq
  c=N+12(\boldsymbol{\rho}\cdot\boldsymbol{\rho})\;.
\eeq
For the case of a single scalar field $N=1$, we obtain the current
\beq\label{T_one_screening}
   T(\varphi)=-\frac{1}{2}(\partial\varphi)^{2}+\left(\frac{1}{\alpha}+\frac{\alpha}{2}\right)\partial^{2}\varphi\;.
\eeq
Note that the same algebra can be defined by the dual screening charge $\oint e^{\alpha^{\vee}\varphi}dz$ with $\alpha^{\vee}=\frac{2}{\alpha}$. In fact, the transformation of $\alpha$ into $\alpha^{\vee}$ can be viewed as the replacement $b \rightarrow b^{-1}$ in Liouville theory, which does not change the central charge. Now let us turn to the general case where there is more than one screening charge and observe that the presence of additional symmetry currents essentially depends on the lengths of the screenings and the scalar products between them.

\subsection{Bosonic and fermionic roots}

\begin{figure}[H]
\begin{center}
\begin{pspicture}(0,0)(2.5,1)
    \pscircle[linewidth=1.5pt](1,0.5){0.2}
    \uput[0](1.5,0.5){$\alpha_r$}
\end{pspicture}
\end{center}
\caption{Representation of a bosonic screening.}
\label{bosonic_root}
\end{figure}
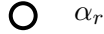

In work \cite{Litvinov:2016mgi}, two types of screening charges (roots) were distinguished, differing by the length value of the vector $\boldsymbol{\alpha}_r$ under consideration. Let us consider these types in more detail. Bosonic roots are depicted as shown in Fig. \ref{bosonic_root}
\noindent and are characterized by the condition that $(\boldsymbol{\alpha}_{r}\cdot \boldsymbol{\alpha}_{r}) \neq -1$. If the current $W_{s}(z)$ satisfies the commutativity conditions \eqref{comm_cond}, then it must have a special form
\beq
  W_{s}(z)=W_{s}\left(T\bigl(\varphi_{\parallel}\bigr),\boldsymbol{\varphi}_{\perp}\right)\;,
\eeq
where
\beq\label{par_orth_fields}
   \varphi_{\parallel}\overset{\text{def}}{=}\frac{(\boldsymbol{\alpha}_{r}\cdot\boldsymbol{\varphi})}{(\boldsymbol{\alpha}_{r}\cdot\boldsymbol{\alpha}_{r})^{\frac{1}{2}}},\quad
   \boldsymbol{\varphi}_{\perp}\overset{\text{def}}{=}\boldsymbol{\varphi}-\frac{(\boldsymbol{\alpha}_{r}\cdot\boldsymbol{\varphi})}{(\boldsymbol{\alpha}_{r}\cdot\boldsymbol{\alpha}_{r})}
   \boldsymbol{\alpha}_{r}\;,
\eeq
and $T\bigl(\varphi_{\parallel}\bigr)$ is given by \eqref{T_one_screening} with $\alpha=(\boldsymbol{\alpha}_{r}\cdot\boldsymbol{\alpha}_{r})^{\frac{1}{2}}$.

Fermionic screenings are depicted as circles with a crosshair in Fig. \ref{fermionic_root}
\begin{figure}[H]
\begin{center}
\begin{pspicture}(0,0)(2.5,1)
    \pscircle[linewidth=1.5pt](1,0.5){0.2}
    \psline[linewidth=1.5pt](0.87,0.37)(1.13,0.63)
    \psline[linewidth=1.5pt](0.87,0.63)(1.13,0.37)
    \uput[0](1.5,0.5){$\alpha_r$}
\end{pspicture}
\end{center}
\caption{Representation of a fermionic screening.}
\label{fermionic_root}
\end{figure}
\noindent where the screening length is $(\boldsymbol{\alpha}_r,\boldsymbol{\alpha}_r)=-1$. In the field space coordinates \eqref{par_orth_fields} defined above, the field $\varphi_{\parallel}$ corresponds to a complex fermion via the Coleman-Mandelstam duality \cite{Coleman:1974bu,Mandelstam:1975hb}. The commutant of the corresponding screening charge $\oint e^{-i\varphi_{\parallel}(z)}dz$ consists of all currents $w_{s}=\psi^{+}\partial^{s-1}\psi$, $s=2,3,\ldots$ Among these currents, only $w_{2}$ and $w_{3}$ are independent, while the rest arise in their operator product expansion. Thus,
\begin{equation}\label{W-for-fermionic-root}
  W_{s}=W_{s}\left(w_{2}\left(\varphi_{\parallel}\right),w_{3}\left(\varphi_{\parallel}\right),\boldsymbol{\varphi}_{\perp}\right)\;.
\end{equation}
Having discussed the constraints imposed on the symmetry currents of the conformal field theory under consideration by bosonic and fermionic screenings, we can proceed to study how they interact with each other and what properties of the considered conformal field theories this leads to.

\subsection{Properties of systems with bosonic and fermionic screenings}

In this section, we list the properties of bosonic and fermionic screening systems known to us, which allow replacing a given set of screenings with an equivalent one, while the symmetry currents corresponding to this set of screenings remain unchanged.

\begin{figure}[H]
\begin{center}
\begin{pspicture}(0,0.5)(3,2.5)
    \pscircle[linewidth=1.5pt](0.7,1){0.2}
    \psline[linewidth=1.5pt](0.57,0.87)(0.83,1.13)
    \psline[linewidth=1.5pt](0.57,1.13)(0.83,0.87)
    \pscircle[linewidth=1.5pt](2.3,1){0.2}
    \psline[linewidth=1.5pt](2.17,0.87)(2.43,1.13)
    \psline[linewidth=1.5pt](2.17,1.13)(2.43,0.87)
    \pscircle[linewidth=1.5pt](1.5,1.8){0.2}
    \psline[linewidth=1.5pt](1.3,2.1)(1.7,2.1)
    \psline[linewidth=1.5pt](0.9,1)(2.1,1)
    \uput[-90](0.7,0.7){$\boldsymbol{\alpha}_1$}
    \uput[-90](2.3,0.7){$\boldsymbol{\alpha}_2$}
    \uput[90](1.5,2.1){$\boldsymbol{\beta}_{12}$}
\end{pspicture}
\end{center}
\caption{Duality between a pair of fermionic screenings and a dressed bosonic screening.}
\label{Wakimoto_bosonic_root_duality}
\end{figure}
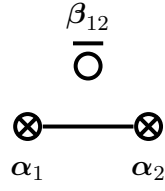

As was shown in \cite{Litvinov:2016mgi}, bosonic roots always appear in pairs
\beq
    \boldsymbol{\alpha}\quad\text{and}\quad \boldsymbol{\alpha}^{\vee}=\frac{2\boldsymbol{\alpha}}{(\boldsymbol{\alpha}\cdot\boldsymbol{\alpha})}\;,
\eeq
meaning that replacing the screening $\boldsymbol{\alpha}$ with $\boldsymbol{\alpha}^{\vee}$ does not change the $W$-algebra under consideration. Another interesting property, this time for fermionic screenings, observed in \cite{Litvinov:2016mgi}, is the duality between dressed/sigma-model Wakimoto-type bosonic screenings \cite{Wakimoto:1986gf} and pairs of fermionic screenings, where $(\boldsymbol{\alpha}_{1}\cdot\boldsymbol{\alpha}_{2})=\xi \neq -1$ are arbitrary:
\beq
   \mathcal{S}_{B}=\oint(\boldsymbol{\alpha}_{1}\cdot\partial\boldsymbol{\varphi})e^{(\boldsymbol{\beta}_{12}\cdot\boldsymbol{\varphi})}dz,\quad\text{where}\quad
   \boldsymbol{\beta}_{12}=\frac{2(\boldsymbol{\alpha}_{1}+\boldsymbol{\alpha}_{2})}{(\boldsymbol{\alpha}_{1}+\boldsymbol{\alpha}_{2})^{2}}\;.
\eeq
This correspondence is illustrated in Fig. \ref{Wakimoto_bosonic_root_duality}

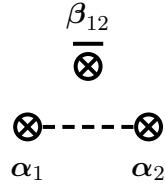
\begin{figure}[H]
\begin{center}
\begin{pspicture}(0,0.5)(3,2.5)
    \pscircle[linewidth=1.5pt](0.7,1){0.2}
    \psline[linewidth=1.5pt](0.57,0.87)(0.83,1.13)
    \psline[linewidth=1.5pt](0.57,1.13)(0.83,0.87)
    \pscircle[linewidth=1.5pt](2.3,1){0.2}
    \psline[linewidth=1.5pt](2.17,0.87)(2.43,1.13)
    \psline[linewidth=1.5pt](2.17,1.13)(2.43,0.87)
    \pscircle[linewidth=1.5pt](1.5,1.8){0.2}
    \psline[linewidth=1.5pt](1.37,1.67)(1.63,1.93)
    \psline[linewidth=1.5pt](1.37,1.93)(1.63,1.67)
    \psline[linewidth=1.5pt](1.3,2.1)(1.7,2.1)
    \psline[linewidth=1.5pt,linestyle=dashed](0.9,1)(2.1,1)
    \uput[-90](0.7,0.7){$\boldsymbol{\alpha}_1$}
    \uput[-90](2.3,0.7){$\boldsymbol{\alpha}_2$}
    \uput[90](1.5,2.1){$\boldsymbol{\beta}_{12}$}
\end{pspicture}
\end{center}
\caption{Duality between a pair of fermionic screenings and a dressed fermionic screening.}
\label{Wakimoto_fermionic_root_duality1}
\end{figure}

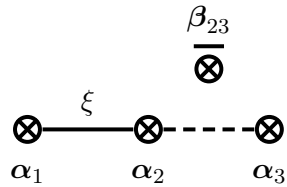
\begin{figure}[H]
\begin{center}
\begin{pspicture}(0,0.5)(4.8,2.5)
    \pscircle[linewidth=1.5pt](0.7,1){0.2}
    \psline[linewidth=1.5pt](0.57,0.87)(0.83,1.13)
    \psline[linewidth=1.5pt](0.57,1.13)(0.83,0.87)
    \pscircle[linewidth=1.5pt](2.3,1){0.2}
    \psline[linewidth=1.5pt](2.17,0.87)(2.43,1.13)
    \psline[linewidth=1.5pt](2.17,1.13)(2.43,0.87)
    \pscircle[linewidth=1.5pt](3.9,1){0.2}
    \psline[linewidth=1.5pt](3.77,0.87)(4.03,1.13)
    \psline[linewidth=1.5pt](3.77,1.13)(4.03,0.87)
    \pscircle[linewidth=1.5pt](3.1,1.8){0.2}
    \psline[linewidth=1.5pt](2.97,1.67)(3.23,1.93)
    \psline[linewidth=1.5pt](2.97,1.93)(3.23,1.67)
    \psline[linewidth=1.5pt](2.9,2.1)(3.3,2.1)
    \psline[linewidth=1.5pt](0.9,1)(2.1,1)
    \psline[linewidth=1.5pt,linestyle=dashed](2.5,1)(3.7,1)
    \uput[-90](0.7,0.7){$\boldsymbol{\alpha}_1$}
    \uput[-90](2.3,0.7){$\boldsymbol{\alpha}_2$}
    \uput[-90](3.9,0.7){$\boldsymbol{\alpha}_3$}
    \uput[90](3.1,2.1){$\boldsymbol{\beta}_{23}$}
    \uput[90](1.5,1){$\xi$}
\end{pspicture}
\end{center}
\caption{Dual dressed fermionic screening embedded in a 3-screening diagram.}
\label{Wakimoto_fermionic_root_duality2}
\end{figure}

Another important type of duality between Wakimoto-type and ordinary screenings is the duality between dressed/sigma-model fermionic screenings and pairs of fermionic screenings with $(\boldsymbol{\alpha}_{1}\cdot\boldsymbol{\alpha}_{2})=-1$:
\beq
   \mathcal{S}_{F}=\oint(\boldsymbol{\alpha}_{1}\cdot\partial\boldsymbol{\varphi})e^{(\boldsymbol{\beta}_{12}\cdot\boldsymbol{\varphi})}dz,\quad\text{where}\quad
   \boldsymbol{\beta}_{12}=\nu\boldsymbol{\alpha}_{1}-(1+\nu)\boldsymbol{\alpha}_{2}
\eeq
The arbitrary parameter $\nu$ here reflects the fact that the Gram matrix of the vectors $\boldsymbol{\alpha}_{1}$ and $\boldsymbol{\alpha}_{2}$ is degenerate. This parameter cannot be fixed if there are only two screenings. Let us depict this situation as follows (here the dashed line corresponds to the scalar product of screening lengths being equal to $-1$) in Fig. \ref{Wakimoto_fermionic_root_duality1}
The subtlety with dressed/sigma-model fermionic screenings is that the parameter $\nu$ cannot be fixed if there are only two screenings $\alpha_1$ and $\alpha_2$. One way to fix the parameter $\nu$ is to consider a larger diagram, for example, the following one in Fig. \ref{Wakimoto_fermionic_root_duality2}
\noindent Then the parameter $\nu$ in the vector $\boldsymbol{\beta}_{23}$ is fixed from the condition
\beq
(\boldsymbol{\beta}_{23}\cdot\boldsymbol{\alpha}_{1})=-1 \quad \implies \quad \nu=-\frac{1}{\xi}\;.
\eeq

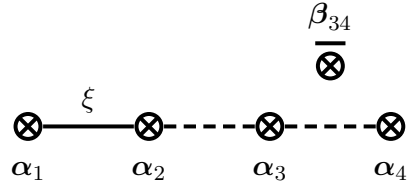
\begin{figure}[H]
\begin{center}
\begin{pspicture}(0,0.5)(4.8,2.5)
    \pscircle[linewidth=1.5pt](-0.1,1){0.2}
    \psline[linewidth=1.5pt](-0.23,0.87)(0.03,1.13)
    \psline[linewidth=1.5pt](-0.23,1.13)(0.03,0.87)
    \pscircle[linewidth=1.5pt](1.5,1){0.2}
    \psline[linewidth=1.5pt](1.37,0.87)(1.63,1.13)
    \psline[linewidth=1.5pt](1.37,1.13)(1.63,0.87)
    \pscircle[linewidth=1.5pt](3.1,1){0.2}
    \psline[linewidth=1.5pt](2.97,0.87)(3.23,1.13)
    \psline[linewidth=1.5pt](2.97,1.13)(3.23,0.87)
    \pscircle[linewidth=1.5pt](4.7,1){0.2}
    \psline[linewidth=1.5pt](4.57,0.87)(4.83,1.13)
    \psline[linewidth=1.5pt](4.57,1.13)(4.83,0.87)
    \pscircle[linewidth=1.5pt](3.9,1.8){0.2}
    \psline[linewidth=1.5pt](3.77,1.67)(4.03,1.93)
    \psline[linewidth=1.5pt](3.77,1.93)(4.03,1.67)
    \psline[linewidth=1.5pt](3.7,2.1)(4.1,2.1)
    \psline[linewidth=1.5pt](0.1,1)(1.3,1)
    \psline[linewidth=1.5pt,linestyle=dashed](1.7,1)(2.9,1)
    \psline[linewidth=1.5pt,linestyle=dashed](3.3,1)(4.5,1)
    \uput[-90](-0.1,0.7){$\boldsymbol{\alpha}_1$}
    \uput[-90](1.5,0.7){$\boldsymbol{\alpha}_2$}
    \uput[-90](3.1,0.7){$\boldsymbol{\alpha}_3$}
    \uput[-90](4.7,0.7){$\boldsymbol{\alpha}_4$}
    \uput[90](3.9,2.1){$\boldsymbol{\beta}_{34}$}
    \uput[90](0.7,1){$\xi$}
\end{pspicture}
\end{center}
\caption{Dual dressed fermionic screening embedded in a 4-screening diagram.}
\label{Wakimoto_fermionic_root_duality3}
\end{figure}

The case in Fig. \ref{Wakimoto_fermionic_root_duality3} is also important for us
Then the parameter $\nu$ in the vector $\boldsymbol{\beta}_{34}$ is fixed from the condition
\beq
   (\boldsymbol{\beta}_{34}\cdot\boldsymbol{\alpha}_{2})=1-\xi\quad\implies\quad
   \nu=\xi-1.
\eeq
Thus, we have examined all cases where a single bosonic screening or a pair of fermionic screenings is dual to a bosonic or a dressed screening of the Wakimoto type, respectively. Now let us proceed to the transformations of screening systems that do not change the corresponding $W$-algebra.

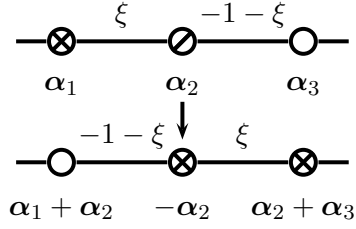
\begin{figure}[H]
\begin{center}
\begin{pspicture}(0,0.5)(4.8,3)
    \pscircle[linewidth=1.5pt](0.7,1){0.2}
    \pscircle[linewidth=1.5pt](2.3,1){0.2}
    \psline[linewidth=1.5pt](2.17,0.87)(2.43,1.13)
    \psline[linewidth=1.5pt](2.17,1.13)(2.43,0.87)
    \pscircle[linewidth=1.5pt](3.9,1){0.2}
    \psline[linewidth=1.5pt](3.77,0.87)(4.03,1.13)
    \psline[linewidth=1.5pt](3.77,1.13)(4.03,0.87)
    \psline[linewidth=1.5pt](0.1,1)(0.5,1)
    \psline[linewidth=1.5pt](0.9,1)(2.1,1)
    \psline[linewidth=1.5pt](2.5,1)(3.7,1)
    \psline[linewidth=1.5pt](4.1,1)(4.5,1)
    \uput[-90](0.7,0.7){$\boldsymbol{\alpha}_1+\boldsymbol{\alpha}_2$}
    \uput[-90](2.3,0.7){$-\boldsymbol{\alpha}_2$}
    \uput[-90](3.9,0.7){$\boldsymbol{\alpha}_2+\boldsymbol{\alpha}_3$}
    \uput[90](1.5,1){$-1-\xi$}
    \uput[90](3.1,1){$\xi$}
    \psline[linewidth=1.5pt]{->}(2.3,1.8)(2.3,1.3)
    \pscircle[linewidth=1.5pt](0.7,2.6){0.2}
    \psline[linewidth=1.5pt](0.57,2.47)(0.83,2.73)
    \psline[linewidth=1.5pt](0.57,2.73)(0.83,2.47)
    \pscircle[linewidth=1.5pt](2.3,2.6){0.2}
    \psline[linewidth=1.5pt](2.17,2.47)(2.43,2.73)
    \psline[linewidth=1.5pt](2.17,2.47)(2.43,2.73)
    \pscircle[linewidth=1.5pt](3.9,2.6){0.2}
    \psline[linewidth=1.5pt](0.1,2.6)(0.5,2.6)
    \psline[linewidth=1.5pt](0.9,2.6)(2.1,2.6)
    \psline[linewidth=1.5pt](2.5,2.6)(3.7,2.6)
    \psline[linewidth=1.5pt](4.1,2.6)(4.5,2.6)
    \uput[-90](0.7,2.3){$\boldsymbol{\alpha}_1$}
    \uput[-90](2.3,2.3){$\boldsymbol{\alpha}_2$}
    \uput[-90](3.9,2.3){$\boldsymbol{\alpha}_3$}
    \uput[90](1.5,2.6){$\xi$}
    \uput[90](3.1,2.6){$-1-\xi$}
\end{pspicture}
\end{center}
\caption{Illustration of fermionic reflection.}
\label{fermionic_reflection_example}
\end{figure}

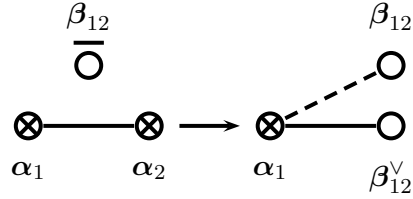
\begin{figure}[H]
\begin{center}
\begin{pspicture}(0,0.5)(7,2.5)
    \pscircle[linewidth=1.5pt](0.7,1){0.2}
    \psline[linewidth=1.5pt](0.57,0.87)(0.83,1.13)
    \psline[linewidth=1.5pt](0.57,1.13)(0.83,0.87)
    \pscircle[linewidth=1.5pt](2.3,1){0.2}
    \psline[linewidth=1.5pt](2.17,0.87)(2.43,1.13)
    \psline[linewidth=1.5pt](2.17,1.13)(2.43,0.87)
    \pscircle[linewidth=1.5pt](1.5,1.8){0.2}
    \psline[linewidth=1.5pt](1.3,2.1)(1.7,2.1)
    \psline[linewidth=1.5pt](0.9,1)(2.1,1)
    \uput[-90](0.7,0.7){$\boldsymbol{\alpha}_1$}
    \uput[-90](2.3,0.7){$\boldsymbol{\alpha}_2$}
    \uput[90](1.5,2.1){$\boldsymbol{\beta}_{12}$}
    \psline[linewidth=1.5pt]{->}(2.7,1)(3.5,1)
    \pscircle[linewidth=1.5pt](3.9,1){0.2}
    \psline[linewidth=1.5pt](3.77,0.87)(4.03,1.13)
    \psline[linewidth=1.5pt](3.77,1.13)(4.03,0.87)
    \pscircle[linewidth=1.5pt](5.5,1){0.2}
    \pscircle[linewidth=1.5pt](5.5,1.8){0.2}
    \psline[linewidth=1.5pt](4.1,1)(5.3,1)
    \psline[linewidth=1.5pt,linestyle=dashed](4.1,1.1)(5.3,1.7)
    \uput[-90](3.9,0.7){$\boldsymbol{\alpha}_1$}
    \uput[-90](5.5,0.7){$\boldsymbol{\beta}_{12}^{\vee}$}
    \uput[90](5.5,2.1){$\boldsymbol{\beta}_{12}$}
\end{pspicture}
\end{center}
\caption{Action of fermionic reflection on a Wakimoto-type screening.}
\label{Wakimoto_bosonic_screening_fermionic_reflection}
\end{figure}

\begin{figure}[H]
\begin{center}
\begin{pspicture}(0,0.5)(10,2.5)
    \pscircle[linewidth=1.5pt](0.7,1){0.2}
    \psline[linewidth=1.5pt](0.57,0.87)(0.83,1.13)
    \psline[linewidth=1.5pt](0.57,1.13)(0.83,0.87)
    \pscircle[linewidth=1.5pt](2.3,1){0.2}
    \psline[linewidth=1.5pt](2.17,0.87)(2.43,1.13)
    \psline[linewidth=1.5pt](2.17,1.13)(2.43,0.87)
    \pscircle[linewidth=1.5pt](3.9,1){0.2}
    \psline[linewidth=1.5pt](3.77,0.87)(4.03,1.13)
    \psline[linewidth=1.5pt](3.77,1.13)(4.03,0.87)
    \pscircle[linewidth=1.5pt](3.1,1.8){0.2}
    \psline[linewidth=1.5pt](2.97,1.67)(3.23,1.93)
    \psline[linewidth=1.5pt](2.97,1.93)(3.23,1.67)
    \psline[linewidth=1.5pt](2.9,2.1)(3.3,2.1)
    \psline[linewidth=1.5pt](0.9,1)(2.1,1)
    \psline[linewidth=1.5pt,linestyle=dashed](2.5,1)(3.7,1)
    \uput[-90](0.7,0.7){$\boldsymbol{\alpha}_1$}
    \uput[-90](2.3,0.7){$\boldsymbol{\alpha}_2$}
    \uput[-90](3.9,0.7){$\boldsymbol{\alpha}_3$}
    \uput[90](3.1,2.1){$\boldsymbol{\beta}_{23}$}
    \uput[90](1.5,1){$\xi$}
    \psline[linewidth=1.5pt]{->}(4.3,1)(5.1,1)
    \pscircle[linewidth=1.5pt](5.5,1){0.2}
    \psline[linewidth=1.5pt](5.37,0.87)(5.63,1.13)
    \psline[linewidth=1.5pt](5.37,1.13)(5.63,0.87)
    \pscircle[linewidth=1.5pt](7.1,1){0.2}
    \psline[linewidth=1.5pt](6.97,0.87)(7.23,1.13)
    \psline[linewidth=1.5pt](6.97,1.13)(7.23,0.87)
    \pscircle[linewidth=1.5pt](8.7,1){0.2}
    \psline[linewidth=1.5pt](8.57,0.87)(8.83,1.13)
    \psline[linewidth=1.5pt](8.57,1.13)(8.83,0.87)
    \pscircle[linewidth=1.5pt](7.9,1.8){0.2}
    \psline[linewidth=1.5pt](7.77,1.67)(8.03,1.93)
    \psline[linewidth=1.5pt](7.77,1.93)(8.03,1.67)
    \psline[linewidth=1.5pt](7.7,2.1)(8.1,2.1)
    \psline[linewidth=1.5pt,linestyle=dashed](5.7,1)(6.9,1)
    \psline[linewidth=1.5pt,linestyle=dashed](7.3,1)(8.5,1)
    \uput[-90](5.5,0.7){$\boldsymbol{\alpha}_1$}
    \uput[-90](7.1,0.7){$\boldsymbol{\beta}_{23}$}
    \uput[-90](8.7,0.7){$-\boldsymbol{\alpha}_3$}
    \uput[90](7.9,2.1){$\boldsymbol{\alpha}_2$}
\end{pspicture}
\end{center}
\caption{Action of fermionic reflection on a Wakimoto-type fermionic screening.}
\label{Wakimoto_fermionic_screening_fermionic_reflection}
\end{figure}
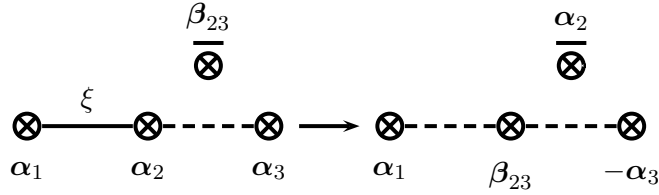

There exists another transformation that affects a given screening and its neighbors, whose scalar product with the screening lengths is non-zero \cite{Litvinov:2016mgi}. This transformation is based on identities for Coulomb integrals derived in work \cite{Baseilhac:1998eq}. If we have a conformal field theory defined by a set of screenings $\mathcal{S}_j=\oint e^{(\boldsymbol{\alpha}_j,\varphi(z))}dz$, $j=1,\ldots,N$, then the same conformal field theory is defined by a set of screenings $\tilde{\mathcal{S}}_j=\oint e^{(\tilde{\boldsymbol{\alpha}}_j,\varphi(z))}dz$ with
\beq
\tilde{\boldsymbol{\alpha}}_j=\begin{cases}
-\boldsymbol{\alpha}_j, \quad \textrm{if} \quad j=r\;, \\
\boldsymbol{\alpha}_j+\boldsymbol{\alpha}_r, \quad \textrm{if} \quad (\boldsymbol{\alpha}_j,\boldsymbol{\alpha}_r) \neq 0\;, \\
\boldsymbol{\alpha}_j \quad \textrm{otherwise},
\end{cases}
\eeq
for the fermionic reflection with respect to the screening $\boldsymbol{\alpha}_r$. This transformation can be demonstrated using the example in Fig. \ref{fermionic_reflection_example}
Note that this transformation can also be interpreted in the case when the screening adjacent to the reflected one is also a fermionic screening, by utilizing a dressed Wakimoto-type screening that is dual to the pair under consideration. This can be best seen using the following diagram in Fig. \ref{Wakimoto_bosonic_screening_fermionic_reflection}
\noindent which demonstrates that under fermionic reflection, the dressed screening $\boldsymbol{\beta}_{12}$ is effectively "undressed" and becomes an ordinary bosonic screening, coexisting in parallel with its dual $\boldsymbol{\beta}_{12}^{\vee}$. Now let us turn to the hitherto unconsidered situation where two fermionic screenings are connected by a scalar product equal to $-1$. To illustrate, let us take the situation where the two fermionic screenings under consideration are embedded in a larger diagram in Fig. \ref{Wakimoto_fermionic_screening_fermionic_reflection}
\noindent from which it follows that under the fermionic reflection of the root $\boldsymbol{\alpha}_3$, the root $\boldsymbol{\beta}_{23}$ is "undressed" and becomes an ordinary fermionic screening, while the root $\boldsymbol{\alpha}_2$, conversely, is "dressed" and becomes a screening dual to the pair of screenings $\boldsymbol{\beta}_{23}$ and $-\boldsymbol{\alpha}_2$. The obtained relations between different types of screenings allow us to proceed to the analysis of conformal field theories defined by various systems of ordinary and Wakimoto-type "dressed" screenings.

Now we possess the necessary tools to analyze screening systems defining conformal field theories whose integrable deformations are the sigma models of interest to us.

\section{Conformal field theories corresponding to deformed \texorpdfstring{$O(N)$}{O(N)} sigma models}

Let us begin formulating our hypothesis with the dual description of $\eta$-deformed $O(N)$ sigma models, for which the case of arbitrary $N$ was studied in detail in \cite{Litvinov:2018bou}. We will focus on odd $N$, since in what follows we will be interested specifically in orthosymplectic models with odd $N$, which were studied by the author in \cite{Alfimov:2020jpy}. To start, let us consider the simplest examples of the $O(3)$ and $O(5)$ models.

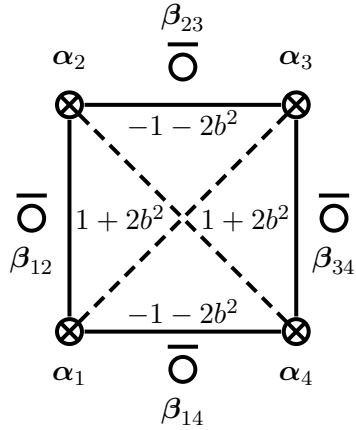
\begin{figure}[H]
\begin{center}
\begin{pspicture}(0,0.2)(5,5.5)
    \pscircle[linewidth=1.5pt](1,1){0.2}
    \psline[linewidth=1.5pt](0.87,0.87)(1.13,1.13)
    \psline[linewidth=1.5pt](0.87,1.13)(1.13,0.87)
    \pscircle[linewidth=1.5pt](4,1){0.2}
    \psline[linewidth=1.5pt](3.87,0.87)(4.13,1.13)
    \psline[linewidth=1.5pt](3.87,1.13)(4.13,0.87)
    \pscircle[linewidth=1.5pt](1,4){0.2}
    \psline[linewidth=1.5pt](0.87,3.87)(1.13,4.13)
    \psline[linewidth=1.5pt](0.87,4.13)(1.13,3.87)
    \pscircle[linewidth=1.5pt](4,4){0.2}
    \psline[linewidth=1.5pt](3.87,3.87)(4.13,4.13)
    \psline[linewidth=1.5pt](3.87,4.13)(4.13,3.87)
    \psline[linewidth=1.5pt](1.2,1)(3.8,1)
    \psline[linewidth=1.5pt](1.2,4)(3.8,4)
    \psline[linewidth=1.5pt](1,1.2)(1,3.8)
    \psline[linewidth=1.5pt](4,1.2)(4,3.8)
    \psline[linewidth=1.5pt,linestyle=dashed](1.13,1.13)(3.87,3.87)
    \psline[linewidth=1.5pt,linestyle=dashed](1.13,3.87)(3.87,1.13)
    \pscircle[linewidth=1.5pt](2.5,0.5){0.2}
    \psline[linewidth=1.5pt](2.3,0.8)(2.7,0.8)
    \pscircle[linewidth=1.5pt](2.5,4.5){0.2}
    \psline[linewidth=1.5pt](2.3,4.8)(2.7,4.8)
    \pscircle[linewidth=1.5pt](0.5,2.5){0.2}
    \psline[linewidth=1.5pt](0.3,2.8)(0.7,2.8)
    \pscircle[linewidth=1.5pt](4.5,2.5){0.2}
    \psline[linewidth=1.5pt](4.3,2.8)(4.7,2.8)
    \uput[-90](1,0.7){$\boldsymbol{\alpha}_1$}
    \uput[90](1,4.3){$\boldsymbol{\alpha}_2$}
    \uput[90](4,4.3){$\boldsymbol{\alpha}_3$}
    \uput[-90](4,0.7){$\boldsymbol{\alpha}_4$}
    \uput[-90](0.5,2.3){$\boldsymbol{\beta}_{12}$}
    \uput[-90](4.5,2.3){$\boldsymbol{\beta}_{34}$}
    \uput[-90](2.5,0.3){$\boldsymbol{\beta}_{14}$}
    \uput[90](2.5,4.8){$\boldsymbol{\beta}_{23}$}
    \uput[0](0.9,2.5){$1+2b^2$}
    \uput[180](4.1,2.5){$1+2b^2$}
    \uput[-90](2.5,4.1){$-1-2b^2$}
    \uput[90](2.5,0.9){$-1-2b^2$}
\end{pspicture}
\end{center}
\caption{The set of fermionic and dressed Wakimoto-type screenings for the $O(3)$ sigma model.}
\label{sausage_model_screenings}
\end{figure}

We would like to verify the compatibility of the metric of a given sigma model with the screening charges defining the $\eta$-deformed $O(3)$ sigma model \cite{Fateev:1992tk}. In work \cite{Alfimov:2021sir}, it was shown that the ultraviolet expansion of both deformations contains dressed Wakimoto-type screenings, which are dual to the ordinary fermionic screenings defining the Toda-type theory for these sigma models. Let us depict these screenings in the diagram in Fig. \ref{sausage_model_screenings}
Using Cartesian coordinates, as in \cite{Litvinov:2018bou}, we can parameterize the lengths of the fermionic screenings as follows
\begin{align}
    & \boldsymbol{\alpha}_1 =bE_1+i\beta e_1\;, \quad \boldsymbol{\alpha}_2 =bE_1-i\beta e_1\;, \\
    & \boldsymbol{\alpha}_3 =-bE_1+i\beta e_1\;, \quad \boldsymbol{\alpha}_4 =-bE_1-i\beta e_1\;, \notag
\end{align}
and these formulas are completely compatible with the scalar products depicted in the diagram above. At the same time, we note that if we take any pair of fermionic screenings from the diagram above, then the entire $\eta$-deformed $O(3)$ sigma model can be viewed as an integrable deformation of a conformal field theory defined, for example, by the screening pairs $\boldsymbol{\alpha}_1$ and $\boldsymbol{\alpha}_2$ or $\boldsymbol{\alpha}_3$ and $\boldsymbol{\alpha}_4$. Moreover, this conformal field theory is nothing other than the sine-Liouville theory, as was shown in works \cite{Fateev:2018yos,Litvinov:2018bou}. Let us now turn to the case $N=5$.

As was shown in \cite{Litvinov:2018bou}, the screening set consists of $5$ fermionic screenings in the case of the $O(5)$ sigma model and, as indicated in the Introduction, the corresponding conformal field theory is based on the coset of affine Lie algebras $\frac{\widehat{\mathfrak{so}}(5)_{-b^2-3}}{\widehat{\mathfrak{so}}(4)_{-b^2-3}}$ with central charge $c=4+\frac{30}{b^2}-\frac{12}{1+b^2}$, leading to the following diagrams in Fig. \ref{O(5)_screening_set} and \ref{so(5)_cft_screening_set} respectively (here $\beta=\sqrt{1+b^2}$):
\begin{figure}[H]
\begin{minipage}{0.45\linewidth}
\begin{center}
\begin{pspicture}(0,0.5)(4,4.5)
    \pscircle[linewidth=1.5pt](1,1){0.2}
    \psline[linewidth=1.5pt](0.87,0.87)(1.13,1.13)
    \psline[linewidth=1.5pt](0.87,1.13)(1.13,0.87)
    \pscircle[linewidth=1.5pt](4,1){0.2}
    \psline[linewidth=1.5pt](3.87,0.87)(4.13,1.13)
    \psline[linewidth=1.5pt](3.87,1.13)(4.13,0.87)
    \pscircle[linewidth=1.5pt](1,4){0.2}
    \psline[linewidth=1.5pt](0.87,3.87)(1.13,4.13)
    \psline[linewidth=1.5pt](0.87,4.13)(1.13,3.87)
    \pscircle[linewidth=1.5pt](4,4){0.2}
    \psline[linewidth=1.5pt](3.87,3.87)(4.13,4.13)
    \psline[linewidth=1.5pt](3.87,4.13)(4.13,3.87)
    \pscircle[linewidth=1.5pt](2.5,2.5){0.2}
    \psline[linewidth=1.5pt](1,1.2)(1,3.8)
    \psline[linewidth=1.5pt](4,1.2)(4,3.8)
    \psline[linewidth=1.5pt](1.13,1.13)(3.87,3.87)
    \psline[linewidth=1.5pt](1.13,3.87)(3.87,1.13)
    \uput[-90](1,0.7){$\boldsymbol{\alpha}_1$}
    \uput[90](1,4.3){$\boldsymbol{\alpha}_2$}
    \uput[-90](2.5,2.3){$\boldsymbol{\alpha}_3$}
    \uput[90](4,4.3){$\boldsymbol{\alpha}_4$}
    \uput[-90](4,0.7){$\boldsymbol{\alpha}_5$}
    \uput[180](1.1,2.5){$1+2b^2$}
    \uput[0](3.9,2.5){$-1-2b^2$}
    \uput[45](1.6,3.2){$-b^2$}
    \uput[-45](1.6,1.8){$-b^2$}
    \uput[135](3.4,3.2){$\beta^2$}
    \uput[-135](3.4,1.8){$\beta^2$}
\end{pspicture}
\end{center}
\caption{The set of fermionic screenings for the $O(5)$ sigma model.}
\label{O(5)_screening_set}
\end{minipage}
\hfill
\begin{minipage}{0.45\linewidth}
\begin{center}
\begin{pspicture}(0,0.5)(4,4.5)
    \pscircle[linewidth=1.5pt](1,1){0.2}
    \psline[linewidth=1.5pt](0.87,0.87)(1.13,1.13)
    \psline[linewidth=1.5pt](0.87,1.13)(1.13,0.87)
    \pscircle[linewidth=1.5pt](4,2.5){0.2}
    \psline[linewidth=1.5pt](3.87,2.37)(4.13,2.63)
    \psline[linewidth=1.5pt](3.87,2.63)(4.13,2.37)
    \pscircle[linewidth=1.5pt](1,4){0.2}
    \psline[linewidth=1.5pt](0.87,3.87)(1.13,4.13)
    \psline[linewidth=1.5pt](0.87,4.13)(1.13,3.87)
    \pscircle[linewidth=1.5pt](2.5,2.5){0.2}
    \psline[linewidth=1.5pt](1,1.2)(1,3.8)
    \psline[linewidth=1.5pt](1.13,1.13)(2.63,2.63)
    \psline[linewidth=1.5pt](1.13,3.87)(2.63,2.37)
    \psline[linewidth=1.5pt](2.7,2.5)(3.8,2.5)
    \uput[-90](1,0.7){$\boldsymbol{\alpha}_1$}
    \uput[90](1,4.3){$\boldsymbol{\alpha}_2$}
    \uput[-90](2.5,2.3){$\boldsymbol{\alpha}_3$}
    \uput[-90](4,2.3){$\boldsymbol{\alpha}_4$}
    \uput[180](1.1,2.5){$1+2b^2$}
    \uput[90](3.2,2.5){$1+b^2$}
    \uput[45](1.6,3.2){$-b^2$}
    \uput[-45](1.6,1.8){$-b^2$}
\end{pspicture}
\end{center}
\caption{The set of fermionic screenings for the $\frac{\widehat{\mathfrak{so}}(5)_{-b^2-3}}{\widehat{\mathfrak{so}}(4)_{-b^2-3}}$ coset conformal field theory.}
\label{so(5)_cft_screening_set}
\end{minipage}
\end{figure}
Applying fermionic reflections to the diagram on the right in various ways, we arrive at the following two diagrams in Fig. \ref{so(5)_cft_first_realization} and \ref{so(5)_cft_second_realization} respectively
\begin{figure}[H]
\begin{minipage}{0.45\linewidth}
\centering
\begin{pspicture}(0,0.5)(7,4.5)
    \pscircle[linewidth=1.5pt](1,2.5){0.2}
    \pscircle[linewidth=1.5pt](2.5,2.5){0.2}
    \pscircle[linewidth=1.5pt](4,2.5){0.2}
    \psline[linewidth=1.5pt](3.87,2.37)(4.13,2.63)
    \psline[linewidth=1.5pt](3.87,2.63)(4.13,2.37)
    \pscircle[linewidth=1.5pt](5.5,2.5){0.2}
	\psline[linewidth=1.5pt,doubleline=true,doublesep=1pt]{<-}(1.2,2.5)(2.3,2.5)
    \psline[linewidth=1.5pt](2.7,2.5)(3.8,2.5)
    \psline[linewidth=1.5pt](4.2,2.5)(5.3,2.5)
    \uput[180](1,2.2){$\boldsymbol{\alpha}_1+\boldsymbol{\alpha}_2$}
    \uput[-90](2.1,2.2){$\boldsymbol{\alpha}_3+\boldsymbol{\alpha}_4$}
    \uput[-90](4.4,2.2){$-\boldsymbol{\alpha}_2-\boldsymbol{\alpha}_3-\boldsymbol{\alpha}_4$}
    \uput[0](5.5,2.2){$\boldsymbol{\alpha}_2+\boldsymbol{\alpha}_3$}
    \uput[90](1.75,2.7){$-2b^2$}
    \uput[90](3.25,2.7){$-b^2$}
    \uput[90](4.75,2.7){$1+b^2$}
\end{pspicture}
\caption{Screening set for the first realization of $\frac{\widehat{\mathfrak{so}}(5)_{-b^2-3}}{\widehat{\mathfrak{so}}(4)_{-b^2-3}}$.}
\label{so(5)_cft_first_realization}
\end{minipage}
\hfill
\begin{minipage}{0.45\linewidth}
\centering
\begin{pspicture}(0,0.5)(6,4.5)
    \pscircle[linewidth=1.5pt](1,1){0.2}
    \pscircle[linewidth=1.5pt](4,2.5){0.2}
    \pscircle[linewidth=1.5pt](1,4){0.2}
    \pscircle[linewidth=1.5pt](2.5,2.5){0.2}
    \psline[linewidth=1.5pt](1.13,1.13)(2.63,2.63)
    \psline[linewidth=1.5pt](1.13,3.87)(2.63,2.37)
    \psline[linewidth=1.5pt](2.7,2.5)(3.8,2.5)
    \uput[-90](1,0.7){$\boldsymbol{\alpha}_1+\boldsymbol{\alpha}_3$}
    \uput[90](1,4.3){$\boldsymbol{\alpha}_2+\boldsymbol{\alpha}_3$}
    \uput[-90](2.5,2.3){$-\boldsymbol{\alpha}_3$}
    \uput[-90](4,2.3){$\boldsymbol{\alpha}_3+\boldsymbol{\alpha}_4$}
    \uput[90](3.2,2.5){$-b^2$}
    \uput[45](1.6,3.2){$1+b^2$}
    \uput[-45](1.6,1.8){$1+b^2$}
\end{pspicture}
\caption{Screening set for the second realization of $\frac{\widehat{\mathfrak{so}}(5)_{-b^2-3}}{\widehat{\mathfrak{so}}(4)_{-b^2-3}}$.}
\label{so(5)_cft_second_realization}
\end{minipage}
\end{figure}

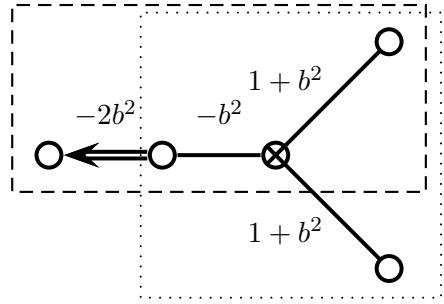
\begin{figure}[H]
\begin{center}
\begin{pspicture}(0,0.5)(7,4.5)
    \pscircle[linewidth=1.5pt](1,2.5){0.2}
    \pscircle[linewidth=1.5pt](2.5,2.5){0.2}
    \pscircle[linewidth=1.5pt](4,2.5){0.2}
    \psline[linewidth=1.5pt](3.87,2.37)(4.13,2.63)
    \psline[linewidth=1.5pt](3.87,2.63)(4.13,2.37)
    \pscircle[linewidth=1.5pt](5.5,4){0.2}
    \pscircle[linewidth=1.5pt](5.5,1){0.2}
	\psline[linewidth=1.5pt,doubleline=true,doublesep=1pt]{<-}(1.2,2.5)(2.3,2.5)
    \psline[linewidth=1.5pt](2.7,2.5)(3.8,2.5)
    \psline[linewidth=1.5pt](4.13,2.63)(5.37,3.87)
    \psline[linewidth=1.5pt](4.13,2.37)(5.37,1.13)
    \uput[90](1.75,2.7){$-2b^2$}
    \uput[90](3.25,2.7){$-b^2$}
    \uput[135](4.75,3.2){$1+b^2$}
    \uput[-135](4.75,1.8){$1+b^2$}
	\psframe[linestyle=dashed](0.5,2)(6,4.5)
	\psframe[linestyle=dotted](2.2,0.6)(6.2,4.4)
\end{pspicture}
\end{center}
\caption{General diagram featuring both realizations of $\frac{\widehat{\mathfrak{so}}(5)_{-b^2-3}}{\widehat{\mathfrak{so}}(4)_{-b^2-3}}$.}
\label{so(5)_cft_both_realizations}
\end{figure}

Let us draw the reader's attention to the fact that both obtained diagrams can be understood as parts of a larger diagram, whose right part looks like the Dynkin diagram for the Lie algebra $\mathfrak{so}(5)$, while its left part looks like the Dynkin diagram of the Lie algebra $\mathfrak{so}(4)$. We illustrate this fact in Fig. \ref{so(5)_cft_both_realizations}
\noindent where the dashed and dotted boxes enclose the screenings corresponding to the two realizations of the $\frac{\widehat{\mathfrak{so}}(5)_{-b^2-3}}{\widehat{\mathfrak{so}}(4)_{-b^2-3}}$ coset conformal field theory, whose deformations define the $\eta$-deformed integrable $O(5)$ sigma model. Based on this observation, we will now formulate a general hypothesis on how to write down, in the general case, the screening set defining the $\frac{\widehat{\mathfrak{so}}(N)_{w}}{\widehat{\mathfrak{so}}(N-1)_{w}}$ conformal field theory whose integrable deformation is the $\eta$-deformed $O(N)$ sigma model.

\begin{figure}[H]
\begin{center}
\begin{pspicture}(0,0.5)(12,4.5)
    \pscircle[linewidth=1.5pt](1,2.5){0.2}
    \psline[linewidth=1.5pt](0.87,2.37)(1.13,2.63)
    \psline[linewidth=1.5pt](0.87,2.63)(1.13,2.37)
    \pscircle[linewidth=1.5pt](2.5,2.5){0.2}
    \psline[linewidth=1.5pt](2.37,2.37)(2.63,2.63)
    \psline[linewidth=1.5pt](2.37,2.63)(2.63,2.37)
    \pscircle[linewidth=1.5pt](4,2.5){0.2}
    \psline[linewidth=1.5pt](3.87,2.37)(4.13,2.63)
    \psline[linewidth=1.5pt](3.87,2.63)(4.13,2.37)
    \pscircle[linewidth=1.5pt](5.5,2.5){0.2}
    \psline[linewidth=1.5pt](5.37,2.37)(5.63,2.63)
    \psline[linewidth=1.5pt](5.37,2.63)(5.63,2.37)
    \pscircle[linewidth=1.5pt](7,2.5){0.2}
    \psline[linewidth=1.5pt](6.87,2.37)(7.13,2.63)
    \psline[linewidth=1.5pt](6.87,2.63)(7.13,2.37)
    \pscircle[linewidth=1.5pt](8.5,2.5){0.2}
    \psline[linewidth=1.5pt](8.37,2.37)(8.63,2.63)
    \psline[linewidth=1.5pt](8.37,2.63)(8.63,2.37)
    \pscircle[linewidth=1.5pt](10,4){0.2}
    \psline[linewidth=1.5pt](9.87,3.87)(10.13,4.13)
    \psline[linewidth=1.5pt](9.87,4.13)(10.13,3.87)
    \pscircle[linewidth=1.5pt](10,1){0.2}
    \psline[linewidth=1.5pt](9.87,0.87)(10.13,1.13)
    \psline[linewidth=1.5pt](9.87,1.13)(10.13,0.87)
	\psline[linewidth=1.5pt](1.2,2.5)(2.3,2.5)
    \psline[linewidth=1.5pt](2.7,2.5)(3.8,2.5)
    \psline[linewidth=1.5pt](4.2,2.5)(4.5,2.5)
    \psline[linewidth=1.5pt,linestyle=dotted](4.5,2.5)(5.0,2.5)
    \psline[linewidth=1.5pt](5.0,2.5)(5.3,2.5)
	\psline[linewidth=1.5pt](5.7,2.5)(6.8,2.5)
    \psline[linewidth=1.5pt](7.2,2.5)(8.3,2.5)
    \psline[linewidth=1.5pt](8.63,2.63)(9.87,3.87)
    \psline[linewidth=1.5pt](8.63,2.37)(9.87,1.13)
    \psline[linewidth=1.5pt](10,3.8)(10,1.2)    
    \uput[90](1.75,2.7){$-b^2$}
    \uput[90](3.25,2.7){$1+b^2$}
    \uput[90](6.25,2.7){$1+b^2$}
    \uput[90](7.75,2.7){$-b^2$}
    \uput[135](9.25,3.2){$1+b^2$}
    \uput[-135](9.25,1.8){$1+b^2$}
    \uput[0](10,2.5){$-1-2b^2$}
\end{pspicture}
\end{center}
\caption{Fermionic realization of the $\frac{\widehat{\mathfrak{so}}(N)_{-b^2-N+2}}{\widehat{\mathfrak{so}}(N-1)_{-b^2-N+2}}$ coset conformal field theory.}
\label{balalaika_diagram}
\end{figure}
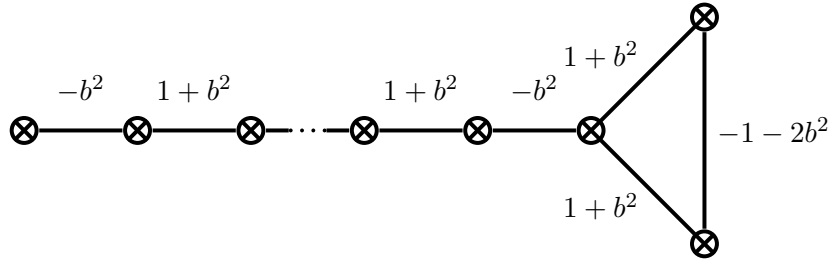

We claim that both realizations of the $\frac{\widehat{\mathfrak{so}}(N)_{w}}{\widehat{\mathfrak{so}}(N-1)_{w}}$ coset conformal field theory by means of screenings can be specified by the screening sets in Fig. \ref{so(N)_cft_both_realizations}, enclosed by the dashed and dotted boxes, respectively:
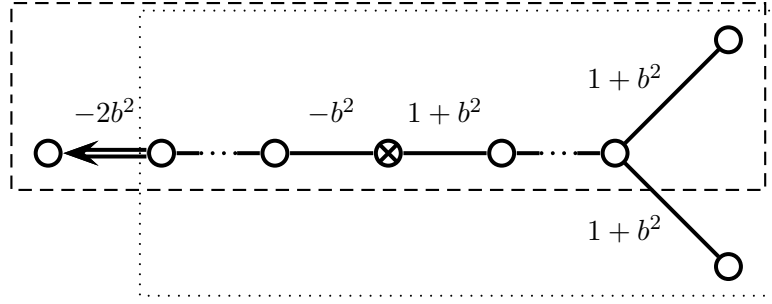
\begin{figure}[H]
\begin{center}
\begin{pspicture}(0,0.5)(12,4.5)
    \pscircle[linewidth=1.5pt](1,2.5){0.2}
    \pscircle[linewidth=1.5pt](2.5,2.5){0.2}
    \pscircle[linewidth=1.5pt](4,2.5){0.2}
    \pscircle[linewidth=1.5pt](5.5,2.5){0.2}
    \pscircle[linewidth=1.5pt](7,2.5){0.2}
    \pscircle[linewidth=1.5pt](8.5,2.5){0.2}
    \psline[linewidth=1.5pt](5.37,2.37)(5.63,2.63)
    \psline[linewidth=1.5pt](5.37,2.63)(5.63,2.37)
    \pscircle[linewidth=1.5pt](10,4){0.2}
    \pscircle[linewidth=1.5pt](10,1){0.2}
	\psline[linewidth=1.5pt,doubleline=true,doublesep=1pt]{<-}(1.2,2.5)(2.3,2.5)
    \psline[linewidth=1.5pt](2.7,2.5)(3.0,2.5)
    \psline[linewidth=1.5pt,linestyle=dotted](3.0,2.5)(3.5,2.5)
    \psline[linewidth=1.5pt](3.5,2.5)(3.8,2.5)
    \psline[linewidth=1.5pt](4.2,2.5)(5.3,2.5)
	\psline[linewidth=1.5pt](5.7,2.5)(6.8,2.5)
    \psline[linewidth=1.5pt](7.2,2.5)(7.5,2.5)
    \psline[linewidth=1.5pt,linestyle=dotted](7.5,2.5)(8.0,2.5)
    \psline[linewidth=1.5pt](8.0,2.5)(8.3,2.5)
    \psline[linewidth=1.5pt](8.63,2.63)(9.87,3.87)
    \psline[linewidth=1.5pt](8.63,2.37)(9.87,1.13)
    \uput[90](1.75,2.7){$-2b^2$}
    \uput[90](4.75,2.7){$-b^2$}
    \uput[90](6.25,2.7){$1+b^2$}
    \uput[135](9.25,3.2){$1+b^2$}
    \uput[-135](9.25,1.8){$1+b^2$}
	\psframe[linestyle=dashed](0.5,2)(10.5,4.5)
	\psframe[linestyle=dotted](2.2,0.6)(10.7,4.4)
\end{pspicture}
\end{center}
\caption{General diagram featuring both realizations of $\frac{\widehat{\mathfrak{so}}(N)_{-b^2-N+2}}{\widehat{\mathfrak{so}}(N-1)_{-b^2-N+2}}$.}
\label{so(N)_cft_both_realizations}
\end{figure}
Applying fermionic reflections in both cases leads to the conformal field theory in Fig. \ref{balalaika_diagram}, whose integrable deformation describes the $\eta$-deformed $O(N)$ sigma model, as was shown in \cite{Litvinov:2018bou}.

Thus, we can state that by using the Dynkin diagrams for the numerator and denominator of the coset conformal field theory, we can generate a screening set for the corresponding coset theory. Let us now turn to the case where the numerator and denominator of the coset are Lie superalgebras, namely the orthosymplectic case.

\section{Conformal field theories corresponding to deformed \texorpdfstring{$OSp(N|2m)$}{OSp(N|2m)} sigma models}

Before proceeding to generalize the formulated hypothesis to the orthosymplectic case, let us remind the reader of the blow-up method, which was used in \cite{Alfimov:2020jpy} to construct a screening system describing the $OSp(N|2m)$ sigma model.

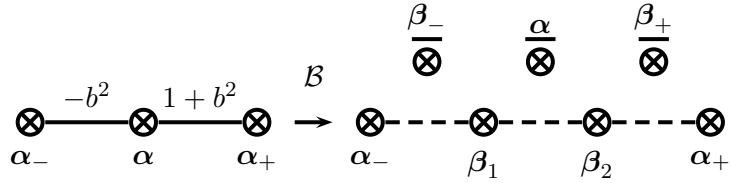
\begin{figure}[H]
\begin{center}
\begin{pspicture}(0,1.5)(12,4)
    \pscircle[linewidth=1.5pt](1,2.5){0.2}
    \psline[linewidth=1.5pt](0.87,2.37)(1.13,2.63)
    \psline[linewidth=1.5pt](0.87,2.63)(1.13,2.37)
    \pscircle[linewidth=1.5pt](2.5,2.5){0.2}
    \psline[linewidth=1.5pt](2.37,2.37)(2.63,2.63)
    \psline[linewidth=1.5pt](2.37,2.63)(2.63,2.37)
    \pscircle[linewidth=1.5pt](4,2.5){0.2}
    \psline[linewidth=1.5pt](3.87,2.37)(4.13,2.63)
    \psline[linewidth=1.5pt](3.87,2.63)(4.13,2.37)
    \pscircle[linewidth=1.5pt](5.5,2.5){0.2}
    \psline[linewidth=1.5pt](5.37,2.37)(5.63,2.63)
    \psline[linewidth=1.5pt](5.37,2.63)(5.63,2.37)
    \pscircle[linewidth=1.5pt](7,2.5){0.2}
    \psline[linewidth=1.5pt](6.87,2.37)(7.13,2.63)
    \psline[linewidth=1.5pt](6.87,2.63)(7.13,2.37)
    \pscircle[linewidth=1.5pt](8.5,2.5){0.2}
    \psline[linewidth=1.5pt](8.37,2.37)(8.63,2.63)
    \psline[linewidth=1.5pt](8.37,2.63)(8.63,2.37)
    \pscircle[linewidth=1.5pt](10,2.5){0.2}
    \psline[linewidth=1.5pt](9.87,2.37)(10.13,2.63)
    \psline[linewidth=1.5pt](9.87,2.63)(10.13,2.37)
    \pscircle[linewidth=1.5pt](6.25,3.3){0.2}
    \psline[linewidth=1.5pt](6.12,3.17)(6.38,3.43)
    \psline[linewidth=1.5pt](6.12,3.43)(6.38,3.17)
    \psline[linewidth=1.5pt](6.05,3.6)(6.45,3.6)
    \pscircle[linewidth=1.5pt](7.75,3.3){0.2}
    \psline[linewidth=1.5pt](7.62,3.17)(7.88,3.43)
    \psline[linewidth=1.5pt](7.62,3.43)(7.88,3.17)
    \psline[linewidth=1.5pt](7.55,3.6)(7.95,3.6)
    \pscircle[linewidth=1.5pt](9.25,3.3){0.2}
    \psline[linewidth=1.5pt](9.12,3.17)(9.38,3.43)
    \psline[linewidth=1.5pt](9.12,3.43)(9.38,3.17)
    \psline[linewidth=1.5pt](9.05,3.6)(9.45,3.6)
	\psline[linewidth=1.5pt](1.2,2.5)(2.3,2.5)
    \psline[linewidth=1.5pt](2.7,2.5)(3.8,2.5)
    \psline[linewidth=1.5pt]{->}(4.5,2.5)(5.0,2.5)
	\psline[linewidth=1.5pt,linestyle=dashed](5.7,2.5)(6.8,2.5)
    \psline[linewidth=1.5pt,linestyle=dashed](7.2,2.5)(8.3,2.5)
    \psline[linewidth=1.5pt,linestyle=dashed](8.7,2.5)(9.8,2.5)  
	\uput[-90](1,2.3){$\boldsymbol{\alpha}_-$}
	\uput[-90](2.5,2.3){$\boldsymbol{\alpha}$}
	\uput[-90](4,2.3){$\boldsymbol{\alpha}_+$}
	\uput[90](4.75,2.8){$\mathcal{B}$}
	\uput[-90](5.5,2.3){$\boldsymbol{\alpha}_-$}
	\uput[-90](7,2.3){$\boldsymbol{\beta}_1$}
	\uput[-90](8.5,2.3){$\boldsymbol{\beta}_2$}
	\uput[-90](10,2.3){$\boldsymbol{\alpha}_+$}
	\uput[90](6.25,3.5){$\boldsymbol{\beta}_-$}
	\uput[90](7.75,3.5){$\boldsymbol{\alpha}$}
	\uput[90](9.25,3.5){$\boldsymbol{\beta}_+$}
    \uput[90](1.75,2.5){$-b^2$}
	\uput[90](3.25,2.5){$1+b^2$}
\end{pspicture}
\end{center}
\caption{Application of the blow-up transformation to a fermionic screening $\boldsymbol{\alpha}$.}
\label{blow-up_transform}
\end{figure}

Let us describe the transformation $\mathcal{B}$ of a root system, which we will call \emph{blow-up}, acting on a sigma model or a conformal field theory as
\beq
O(N)\rightarrow OSp(N|2)\;,
\eeq
or, more generally, as
\beq
OSp(N|2m)\rightarrow OSp(N|2m+2)\;.
\eeq
It can be applied both to the diagram corresponding to a conformal field theory and to its integrable deformation. Let us consider the case of odd $N=2n+1$. On a fermionic root $\boldsymbol{\alpha}$, it acts as follows
\beq
   \boldsymbol{\alpha}=-b\boldsymbol{E}+i\beta\boldsymbol{e}\overset{\mathcal{B}}{\longrightarrow} \{\boldsymbol{\beta}_{1},\boldsymbol{\beta}_{2}\}=
   \left\{-\frac{1}{b}\boldsymbol{E}+\frac{i\beta}{b}\boldsymbol{\epsilon},\frac{ib}{\beta}\boldsymbol{\epsilon}-\frac{i}{\beta}\boldsymbol{e}\right\}\;,
\eeq
where $\boldsymbol{\epsilon}$ is a new basis vector. In the case where the root being blown up is located inside the diagram, this can be depicted as shown in Fig. \ref{blow-up_transform}
Let us apply this to the case of screening charges of the deformed $OSp(N|2)$ sigma model. Consider the simplest case of the $OSp(5|2)$ diagram in Fig. \ref{OSp(5|2)_screening_set}. According to our rule, it is obtained from the $O(5)$ diagram in Fig. \ref{O(5)_Toda_screening_set} by blowing up the root $\boldsymbol{\alpha}_{3}$:
\begin{figure}[H]
\begin{minipage}{0.4\linewidth}
\centering
\begin{pspicture}(0,0.2)(6,5)
    \pscircle[linewidth=1.5pt](1,1){0.2}
    \psline[linewidth=1.5pt](0.87,0.87)(1.13,1.13)
    \psline[linewidth=1.5pt](0.87,1.13)(1.13,0.87)
    \pscircle[linewidth=1.5pt](4,1){0.2}
    \psline[linewidth=1.5pt](3.87,0.87)(4.13,1.13)
    \psline[linewidth=1.5pt](3.87,1.13)(4.13,0.87)
    \pscircle[linewidth=1.5pt](1,4){0.2}
    \psline[linewidth=1.5pt](0.87,3.87)(1.13,4.13)
    \psline[linewidth=1.5pt](0.87,4.13)(1.13,3.87)
    \pscircle[linewidth=1.5pt](4,4){0.2}
    \psline[linewidth=1.5pt](3.87,3.87)(4.13,4.13)
    \psline[linewidth=1.5pt](3.87,4.13)(4.13,3.87)
    \pscircle[linewidth=1.5pt](2.5,2.5){0.2}
    \psline[linewidth=1.5pt](1,1.2)(1,3.8)
    \psline[linewidth=1.5pt](4,1.2)(4,3.8)
    \psline[linewidth=1.5pt](1.13,1.13)(3.87,3.87)
    \psline[linewidth=1.5pt](1.13,3.87)(3.87,1.13)
    \uput[-90](1,0.7){$\boldsymbol{\alpha}_1$}
    \uput[90](1,4.3){$\boldsymbol{\alpha}_2$}
    \uput[-90](2.5,2.3){$\boldsymbol{\alpha}_3$}
    \uput[90](4,4.3){$\boldsymbol{\alpha}_4$}
    \uput[-90](4,0.7){$\boldsymbol{\alpha}_5$}
    \uput[180](1.1,2.5){$1+2b^2$}
    \uput[0](3.9,2.5){$-1-2b^2$}
    \uput[45](1.6,3.2){$-b^2$}
    \uput[-45](1.6,1.8){$-b^2$}
    \uput[135](3.4,3.2){$\beta^2$}
    \uput[-135](3.4,1.8){$\beta^2$}
\end{pspicture}
\caption{The set of fermionic screenings for the $\eta$-deformed $O(5)$ sigma model.}
\label{O(5)_Toda_screening_set}
\end{minipage}
\hfill
\begin{minipage}{0.4\linewidth}
\centering
\begin{pspicture}(0,0.2)(8,5)
    \pscircle[linewidth=1.5pt](1,1){0.2}
    \psline[linewidth=1.5pt](0.87,0.87)(1.13,1.13)
    \psline[linewidth=1.5pt](0.87,1.13)(1.13,0.87)
    \pscircle[linewidth=1.5pt](5.5,1){0.2}
    \psline[linewidth=1.5pt](5.37,0.87)(5.63,1.13)
    \psline[linewidth=1.5pt](5.37,1.13)(5.63,0.87)
    \pscircle[linewidth=1.5pt](1,4){0.2}
    \psline[linewidth=1.5pt](0.87,3.87)(1.13,4.13)
    \psline[linewidth=1.5pt](0.87,4.13)(1.13,3.87)
    \pscircle[linewidth=1.5pt](5.5,4){0.2}
    \psline[linewidth=1.5pt](5.37,3.87)(5.63,4.13)
    \psline[linewidth=1.5pt](5.37,4.13)(5.63,3.87)
    \pscircle[linewidth=1.5pt](2.5,2.5){0.2}
    \psline[linewidth=1.5pt](2.37,2.37)(2.63,2.63)
    \psline[linewidth=1.5pt](2.37,2.63)(2.63,2.37)
    \pscircle[linewidth=1.5pt](4,2.5){0.2}
    \psline[linewidth=1.5pt](3.87,2.37)(4.13,2.63)
    \psline[linewidth=1.5pt](3.87,2.63)(4.13,2.37)
    \pscircle[linewidth=1.5pt](3.25,3.3){0.2}
    \psline[linewidth=1.5pt](3.12,3.17)(3.38,3.43)
    \psline[linewidth=1.5pt](3.12,3.43)(3.38,3.17)
    \psline[linewidth=1.5pt](3.15,3.6)(3.35,3.6)
    \pscircle[linewidth=1.5pt](0.2,2.5){0.2}
    \psline[linewidth=1.5pt](0.1,2.8)(0.3,2.8)
    \pscircle[linewidth=1.5pt](6.3,2.5){0.2}
    \psline[linewidth=1.5pt](6.2,2.8)(6.4,2.8)
    \pscircle[linewidth=1.5pt](2.2,3.6){0.2}
    \psline[linewidth=1.5pt](2.07,3.47)(2.33,3.73)
    \psline[linewidth=1.5pt](2.07,3.73)(2.33,3.47)
    \psline[linewidth=1.5pt](2.1,3.9)(2.3,3.9)
    \pscircle[linewidth=1.5pt](2.2,1.4){0.2}
    \psline[linewidth=1.5pt](2.07,1.27)(2.33,1.53)
    \psline[linewidth=1.5pt](2.07,1.53)(2.33,1.27)
    \psline[linewidth=1.5pt](2.1,1.7)(2.3,1.7)
    \pscircle[linewidth=1.5pt](4.2,3.6){0.2}
    \psline[linewidth=1.5pt](4.07,3.47)(4.33,3.73)
    \psline[linewidth=1.5pt](4.07,3.73)(4.33,3.47)
    \psline[linewidth=1.5pt](4.1,3.9)(4.3,3.9)
    \pscircle[linewidth=1.5pt](4.2,1.4){0.2}
    \psline[linewidth=1.5pt](4.07,1.27)(4.33,1.53)
    \psline[linewidth=1.5pt](4.07,1.53)(4.33,1.27)
    \psline[linewidth=1.5pt](4.1,1.7)(4.3,1.7)
    \psline[linewidth=1.5pt](1,1.2)(1,3.8)
    \psline[linewidth=1.5pt](5.5,1.2)(5.5,3.8)
    \psline[linewidth=1.5pt,linestyle=dashed](1.13,1.13)(2.37,2.37)
    \psline[linewidth=1.5pt,linestyle=dashed](1.13,3.87)(2.37,2.63)
    \psline[linewidth=1.5pt,linestyle=dashed](4.13,2.63)(5.37,3.87)
    \psline[linewidth=1.5pt,linestyle=dashed](4.13,2.37)(5.37,1.13)
    \psline[linewidth=1.5pt,linestyle=dashed](2.7,2.5)(3.8,2.5)
    \uput[-90](1,0.8){$\boldsymbol{\alpha}_1$}
    \uput[90](1,4.2){$\boldsymbol{\alpha}_2$}
    \uput[-90](2.5,2.3){$\boldsymbol{\beta}_1$}
    \uput[90](3.25,3.5){$\boldsymbol{\alpha}_3$}
    \uput[-90](4,2.3){$\boldsymbol{\beta}_2$}
    \uput[90](5.5,4.2){$\boldsymbol{\alpha}_4$}
    \uput[-90](5.5,0.8){$\boldsymbol{\alpha}_5$}
    \uput[-90](0.2,2.3){$\boldsymbol{\beta}_{12}$}
    \uput[-90](6.3,2.3){$\boldsymbol{\beta}_{45}$}
    \uput[-90](2.2,1.3){$\boldsymbol{\beta}_-^-$}
    \uput[-90](4.2,1.3){$\boldsymbol{\beta}_+^-$}
    \uput[90](2.2,3.8){$\boldsymbol{\beta}_-^+$}
    \uput[90](4.2,3.8){$\boldsymbol{\beta}_+^+$}
\end{pspicture}
\caption{The set of fermionic screenings for the $\eta$-deformed $OSp(5|2)$ sigma model.}
\label{OSp(5|2)_screening_set}
\end{minipage}
\end{figure}

The vectors $\boldsymbol{\alpha}_{r}$ can be parameterized as follows ($\beta=\sqrt{1+b^{2}}$):
\begin{align}
& \boldsymbol{\alpha}_{1}=b\boldsymbol{E}_{1}+i\beta\boldsymbol{e}_{1}\;, \quad \boldsymbol{\alpha}_{2}=b\boldsymbol{E}_{1}-i\beta\boldsymbol{e}_{1}\;, \quad \boldsymbol{\alpha}_{3}=-b\boldsymbol{E}_{1}+i\beta\boldsymbol{e}_{2}\;, \\
& \boldsymbol{\alpha}_{4}=b\boldsymbol{E}_{2}-i\beta\boldsymbol{e}_{2}\;, \quad \boldsymbol{\alpha}_{5}=-b\boldsymbol{E}_{2}-i\beta\boldsymbol{e}_{2}\;, \notag \\
& \boldsymbol{\beta}_{1}=-\frac{1}{b}\boldsymbol{E}_{1}+\frac{i\beta}{b}\boldsymbol{\epsilon}\;, \quad \boldsymbol{\beta}_{2}=\frac{ib}{\beta}\boldsymbol{\epsilon}-\frac{i}{\beta}\boldsymbol{e}_{2}\;, \quad \boldsymbol{\beta}_{-}^{\pm}=\pm\frac{i}{\beta}\boldsymbol{e}_{1}-\frac{ib}{\beta}\boldsymbol{\epsilon}\;, \notag \\
& \boldsymbol{\beta}_{+}^{\pm}=\pm\frac{1}{b}\boldsymbol{E}_{2}-\frac{i\beta}{b}\boldsymbol{\epsilon}\;, \quad \boldsymbol{\beta}_{12}=\frac{1}{b}\boldsymbol{E}_{1}\;, \quad \boldsymbol{\beta}_{45}=\frac{i}{\beta}\boldsymbol{e}_{2}\;. \notag
\end{align}

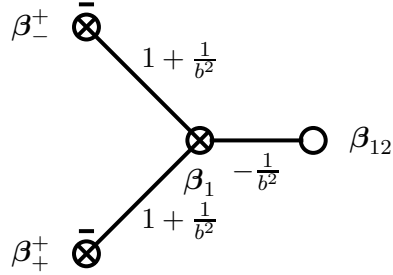
\begin{figure}[H]
\begin{center}
\begin{pspicture}(0,0.5)(5,4.5)
    \pscircle[linewidth=1.5pt](1,1){0.2}
    \psline[linewidth=1.5pt](0.87,0.87)(1.13,1.13)
    \psline[linewidth=1.5pt](0.87,1.13)(1.13,0.87)
    \psline[linewidth=1.5pt](0.9,1.3)(1.1,1.3)
    \pscircle[linewidth=1.5pt](1,4){0.2}
    \psline[linewidth=1.5pt](0.87,3.87)(1.13,4.13)
    \psline[linewidth=1.5pt](0.87,4.13)(1.13,3.87)
    \psline[linewidth=1.5pt](0.9,4.3)(1.1,4.3)
    \pscircle[linewidth=1.5pt](4,2.5){0.2}
    \pscircle[linewidth=1.5pt](2.5,2.5){0.2}
    \psline[linewidth=1.5pt](1.13,1.13)(2.63,2.63)
    \psline[linewidth=1.5pt](1.13,3.87)(2.63,2.37)
    \psline[linewidth=1.5pt](2.63,2.5)(3.87,2.5)
    \uput[180](0.7,1){$\boldsymbol{\beta}_+^+$}
    \uput[180](0.7,4){$\boldsymbol{\beta}_-^+$}
    \uput[-90](2.5,2.3){$\boldsymbol{\beta}_1$}
    \uput[0](4.3,2.5){$\boldsymbol{\beta}_{12}$}
    \uput[-90](3.25,2.5){$-\frac{1}{b^2}$}
    \uput[45](1.6,3.2){$1+\frac{1}{b^2}$}
    \uput[-45](1.6,1.8){$1+\frac{1}{b^2}$}
\end{pspicture}
\end{center}
\caption{The set of screenings arising in the ultraviolet limit of the $OSp(5|2)$ sigma model.}
\label{OSp(5|2)_sigma_model_screening_set}
\end{figure}

Let us now turn to specifying the metric of the deformed $OSp(5|2)$ sigma model. Taking the dual screenings, we obtain the following system in Fig. \ref{OSp(5|2)_sigma_model_screening_set}, which includes dressed Wakimoto-type screenings.
Choosing $z=x^1-ix^2$ ($\bar{z}=x^1+ix^2$) and then performing the Wick rotation $x^2=ix^0$, we obtain the action in the Minkowski signature:
\begin{multline}
\mathcal{L}=\frac{1}{8\pi}\left(\sum\limits_{i=1}^2 (\partial_+ \Phi_i)(\partial_- \Phi_i)+\sum\limits_{j=1}^3 (\partial_+ \phi_j)(\partial_- \phi_j)\right)+ \\
+\Lambda_1 e^{-\frac{i\beta}{b}\phi_3}\left(\partial_+ \left(b\Phi_2+i\beta\phi_2\right)\partial_- \left(b\Phi_2-i\beta\phi_2\right)e^{-\frac{\Phi_2}{b}}+\right. \\
\left.+\partial_+ \left(b\Phi_2-i\beta\phi_2\right)\partial_- \left(b\Phi_2+i\beta\phi_2\right)e^{\frac{\Phi_2}{b}}\right)+\Lambda_2 e^{-\frac{\Phi_1}{b}+\frac{i\beta}{b}\phi_3}+ \\
+\Lambda_3 \partial_+ \left(b\Phi_1+i\beta\phi_1\right)\partial_- \left(b\Phi_1-i\beta\phi_1\right)e^{\frac{\Phi_1}{b}}+\frac{\pi b^2}{\beta^2}\Lambda_1 \Lambda_2 e^{\frac{\Phi_1}{b}}\times \\
\times \left(\partial_+ \left(b\Phi_2+i\beta\phi_2\right)\partial_- \left(b\Phi_2-i\beta\phi_2\right)e^{-\frac{\Phi_2}{b}}+\partial_+ \left(b\Phi_2-i\beta\phi_2\right)\partial_- \left(b\Phi_2+i\beta\phi_2\right)e^{\frac{\Phi_2}{b}}\right)+\ldots\;,
\end{multline}
Fermionizing the field $\varphi_3$ and integrating over half of the fermion components, we recover the ultraviolet limit of the one-loop action of the sigma model after rescaling the field by $b$.

Screening charges can also be written out for the limit $b \rightarrow 0$. Taking the subsystem of screenings in Fig. \ref{OSp(5|2)_Toda_screening_set} that are regular in the limit $b \rightarrow 0$:
\begin{figure}[H]
\begin{center}
\begin{pspicture}(0,0.5)(5,4.5)
    \pscircle[linewidth=1.5pt](1,1){0.2}
    \psline[linewidth=1.5pt](0.87,0.87)(1.13,1.13)
    \psline[linewidth=1.5pt](0.87,1.13)(1.13,0.87)
    \pscircle[linewidth=1.5pt](4,1){0.2}
    \psline[linewidth=1.5pt](3.87,0.87)(4.13,1.13)
    \psline[linewidth=1.5pt](3.87,1.13)(4.13,0.87)
    \pscircle[linewidth=1.5pt](1,4){0.2}
    \psline[linewidth=1.5pt](0.87,3.87)(1.13,4.13)
    \psline[linewidth=1.5pt](0.87,4.13)(1.13,3.87)
    \pscircle[linewidth=1.5pt](4,4){0.2}
    \psline[linewidth=1.5pt](3.87,3.87)(4.13,4.13)
    \psline[linewidth=1.5pt](3.87,4.13)(4.13,3.87)
    \pscircle[linewidth=1.5pt](2.5,2.5){0.2}
    \psline[linewidth=1.5pt](2.4,2.8)(2.6,2.8)
    \psline[linewidth=1.5pt](1,1.2)(1,3.8)
    \psline[linewidth=1.5pt](4,1.2)(4,3.8)
    \psline[linewidth=1.5pt](1.13,1.13)(3.87,3.87)
    \psline[linewidth=1.5pt](1.13,3.87)(3.87,1.13)
    \uput[-90](1,0.7){$\boldsymbol{\alpha}_1$}
    \uput[90](1,4.3){$\boldsymbol{\alpha}_2$}
    \uput[-90](2.5,2.3){$\boldsymbol{\alpha}_3$}
    \uput[90](4,4.3){$\boldsymbol{\alpha}_4$}
    \uput[-90](4,0.7){$\boldsymbol{\alpha}_5$}
    \uput[180](1.1,2.5){$1+2b^2$}
    \uput[0](3.9,2.5){$-1-2b^2$}
    \uput[45](1.6,3.2){$-b^2$}
    \uput[-45](1.6,1.8){$-b^2$}
    \uput[135](3.4,3.2){$\beta^2$}
    \uput[-135](3.4,1.8){$\beta^2$}
\end{pspicture}
\end{center}
\caption{The set of screenings for the dual description of the $\eta$-deformed $OSp(5|2)$ sigma model in the limit $b \rightarrow 0$.}
\label{OSp(5|2)_Toda_screening_set}
\end{figure}
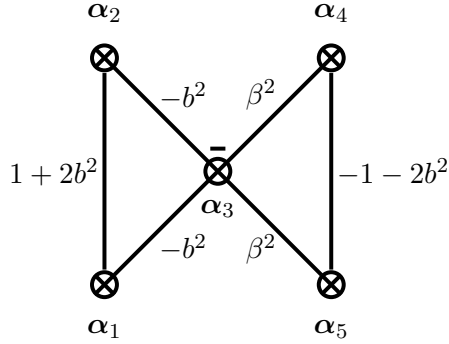
we can write the Lagrangian of the dual model:
\begin{multline}
\mathcal{L}=\frac{1}{8\pi}\left(\sum\limits_{i=1}^2 (\partial \Phi_i)(\bar{\partial} \Phi_i)+\sum\limits_{j=1}^3 (\partial \varphi_j)(\bar{\partial} \varphi_j)\right)+2\Lambda_1 e^{b\Phi_1}\cos \beta \varphi_1+ \\
+\Lambda_2\partial(\Phi_1-i\beta\varphi_3)\bar{\partial}(\Phi_1+i\beta\varphi_3)e^{-b\Phi_1+i\beta\varphi_2}+\\
+\Lambda_3\left(e^{-b\Phi_2-i\beta\varphi_2}+e^{b\Phi_2-i\beta\varphi_2}\right)+(\textrm{counterterms})\;.
\end{multline}
This action has a finite number of counterterms! Concluding our discussion of the $OSp(5|2)$ case, we note that, by analogy with the case of the $O(5)$ sigma model, from the affine diagram we can extract the diagram in Fig. \ref{osp(5|2)_cft_screeinig_set} corresponding to the conformal field theory whose integrable perturbation yields the deformed $OSp(5|2)$ sigma model:
\begin{figure}[H]
\begin{center}
\begin{pspicture}(0,0.5)(7,4.5)
    \pscircle[linewidth=1.5pt](1,1){0.2}
    \psline[linewidth=1.5pt](0.87,0.87)(1.13,1.13)
    \psline[linewidth=1.5pt](0.87,1.13)(1.13,0.87)
    \pscircle[linewidth=1.5pt](1,4){0.2}
    \psline[linewidth=1.5pt](0.87,3.87)(1.13,4.13)
    \psline[linewidth=1.5pt](0.87,4.13)(1.13,3.87)
    \pscircle[linewidth=1.5pt](5.5,2.5){0.2}
    \psline[linewidth=1.5pt](5.37,2.37)(5.63,2.63)
    \psline[linewidth=1.5pt](5.37,2.63)(5.63,2.37)
    \pscircle[linewidth=1.5pt](2.5,2.5){0.2}
    \psline[linewidth=1.5pt](2.37,2.37)(2.63,2.63)
    \psline[linewidth=1.5pt](2.37,2.63)(2.63,2.37)
    \pscircle[linewidth=1.5pt](4,2.5){0.2}
    \psline[linewidth=1.5pt](3.87,2.37)(4.13,2.63)
    \psline[linewidth=1.5pt](3.87,2.63)(4.13,2.37)
    \psline[linewidth=1.5pt](1,1.2)(1,3.8)
    \psline[linewidth=1.5pt,linestyle=dashed](1.13,1.13)(2.37,2.37)
    \psline[linewidth=1.5pt,linestyle=dashed](1.13,3.87)(2.37,2.63)
    \psline[linewidth=1.5pt,linestyle=dashed](2.7,2.5)(3.8,2.5)
    \psline[linewidth=1.5pt,linestyle=dashed](4.2,2.5)(5.3,2.5)
    \uput[-90](1,0.8){$\boldsymbol{\alpha}_1$}
    \uput[90](1,4.2){$\boldsymbol{\alpha}_2$}
    \uput[-90](2.5,2.3){$\boldsymbol{\beta}_1$}
    \uput[-90](4,2.3){$\boldsymbol{\beta}_2$}
    \uput[-90](5.5,2.3){$\boldsymbol{\alpha}_4$}
    \uput[180](1,2.5){$1+2b^2$}
\end{pspicture}
\end{center}
\caption{The set of screenings for the $\frac{\widehat{\mathfrak{osp}}(5|2)_w}{\widehat{\mathfrak{osp}}(4|2)_w}$ coset conformal field theory.}
\label{osp(5|2)_cft_screeinig_set}
\end{figure}
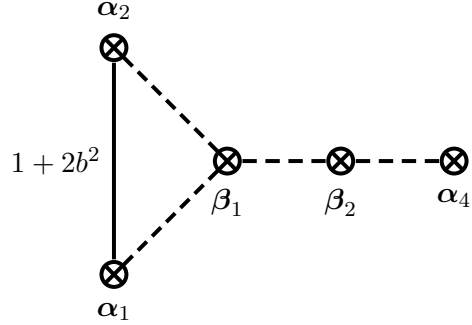

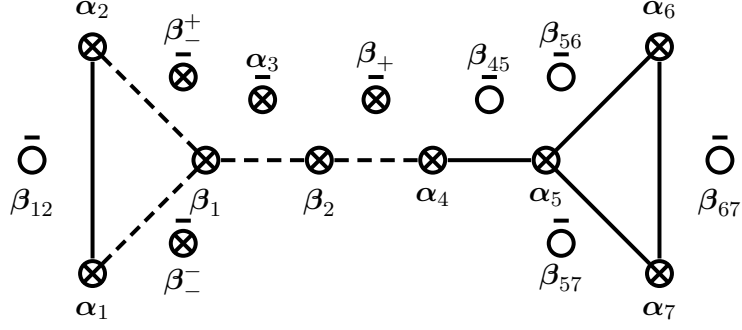
\begin{figure}[H]
\begin{center}
\begin{pspicture}(0,0.5)(9.5,4.5)
    \pscircle[linewidth=1.5pt](1,1){0.2}
    \psline[linewidth=1.5pt](0.87,0.87)(1.13,1.13)
    \psline[linewidth=1.5pt](0.87,1.13)(1.13,0.87)
    \pscircle[linewidth=1.5pt](8.5,1){0.2}
    \psline[linewidth=1.5pt](8.37,0.87)(8.63,1.13)
    \psline[linewidth=1.5pt](8.37,1.13)(8.63,0.87)
    \pscircle[linewidth=1.5pt](1,4){0.2}
    \psline[linewidth=1.5pt](0.87,3.87)(1.13,4.13)
    \psline[linewidth=1.5pt](0.87,4.13)(1.13,3.87)
    \pscircle[linewidth=1.5pt](8.5,4){0.2}
    \psline[linewidth=1.5pt](8.37,3.87)(8.63,4.13)
    \psline[linewidth=1.5pt](8.37,4.13)(8.63,3.87)
    \pscircle[linewidth=1.5pt](2.5,2.5){0.2}
    \psline[linewidth=1.5pt](2.37,2.37)(2.63,2.63)
    \psline[linewidth=1.5pt](2.37,2.63)(2.63,2.37)
    \pscircle[linewidth=1.5pt](4,2.5){0.2}
    \psline[linewidth=1.5pt](3.87,2.37)(4.13,2.63)
    \psline[linewidth=1.5pt](3.87,2.63)(4.13,2.37)
    \pscircle[linewidth=1.5pt](5.5,2.5){0.2}
    \psline[linewidth=1.5pt](5.37,2.37)(5.63,2.63)
    \psline[linewidth=1.5pt](5.37,2.63)(5.63,2.37)
    \pscircle[linewidth=1.5pt](7,2.5){0.2}
    \psline[linewidth=1.5pt](6.87,2.37)(7.13,2.63)
    \psline[linewidth=1.5pt](6.87,2.63)(7.13,2.37)
    \pscircle[linewidth=1.5pt](3.25,3.3){0.2}
    \psline[linewidth=1.5pt](3.12,3.17)(3.38,3.43)
    \psline[linewidth=1.5pt](3.12,3.43)(3.38,3.17)
    \psline[linewidth=1.5pt](3.15,3.6)(3.35,3.6)
    \pscircle[linewidth=1.5pt](0.2,2.5){0.2}
    \psline[linewidth=1.5pt](0.1,2.8)(0.3,2.8)
    \pscircle[linewidth=1.5pt](9.3,2.5){0.2}
    \psline[linewidth=1.5pt](9.2,2.8)(9.4,2.8)
    \pscircle[linewidth=1.5pt](2.2,3.6){0.2}
    \psline[linewidth=1.5pt](2.07,3.47)(2.33,3.73)
    \psline[linewidth=1.5pt](2.07,3.73)(2.33,3.47)
    \psline[linewidth=1.5pt](2.1,3.9)(2.3,3.9)
    \pscircle[linewidth=1.5pt](2.2,1.4){0.2}
    \psline[linewidth=1.5pt](2.07,1.27)(2.33,1.53)
    \psline[linewidth=1.5pt](2.07,1.53)(2.33,1.27)
    \psline[linewidth=1.5pt](2.1,1.7)(2.3,1.7)
    \pscircle[linewidth=1.5pt](7.2,3.6){0.2}
    \psline[linewidth=1.5pt](7.1,3.9)(7.3,3.9)
    \pscircle[linewidth=1.5pt](7.2,1.4){0.2}
    \psline[linewidth=1.5pt](7.1,1.7)(7.3,1.7)
    \pscircle[linewidth=1.5pt](4.75,3.3){0.2}
    \psline[linewidth=1.5pt](4.62,3.17)(4.88,3.43)
    \psline[linewidth=1.5pt](4.62,3.43)(4.88,3.17)
    \psline[linewidth=1.5pt](4.65,3.6)(4.85,3.6)
    \pscircle[linewidth=1.5pt](6.25,3.3){0.2}
    \psline[linewidth=1.5pt](6.15,3.6)(6.35,3.6)
    \psline[linewidth=1.5pt](1,1.2)(1,3.8)
    \psline[linewidth=1.5pt](8.5,1.2)(8.5,3.8)
    \psline[linewidth=1.5pt,linestyle=dashed](1.13,1.13)(2.37,2.37)
    \psline[linewidth=1.5pt,linestyle=dashed](1.13,3.87)(2.37,2.63)
    \psline[linewidth=1.5pt](7.13,2.63)(8.37,3.87)
    \psline[linewidth=1.5pt](7.13,2.37)(8.37,1.13)
    \psline[linewidth=1.5pt,linestyle=dashed](2.7,2.5)(3.8,2.5)
    \psline[linewidth=1.5pt,linestyle=dashed](4.2,2.5)(5.3,2.5)
    \psline[linewidth=1.5pt](5.7,2.5)(6.8,2.5)
    \uput[-90](1,0.8){$\boldsymbol{\alpha}_1$}
    \uput[90](1,4.2){$\boldsymbol{\alpha}_2$}
    \uput[-90](2.5,2.3){$\boldsymbol{\beta}_1$}
    \uput[90](3.25,3.5){$\boldsymbol{\alpha}_3$}
    \uput[-90](4,2.3){$\boldsymbol{\beta}_2$}
    \uput[90](4.75,3.5){$\boldsymbol{\beta}_+$}
    \uput[-90](5.5,2.3){$\boldsymbol{\alpha}_4$}
    \uput[90](6.25,3.5){$\boldsymbol{\beta}_{45}$}
    \uput[-90](7,2.3){$\boldsymbol{\alpha}_5$}
    \uput[90](8.5,4.2){$\boldsymbol{\alpha}_6$}
    \uput[-90](8.5,0.8){$\boldsymbol{\alpha}_7$}
    \uput[-90](0.2,2.3){$\boldsymbol{\beta}_{12}$}
    \uput[-90](9.3,2.3){$\boldsymbol{\beta}_{67}$}
    \uput[-90](2.2,1.3){$\boldsymbol{\beta}_-^-$}
    \uput[-90](7.2,1.3){$\boldsymbol{\beta}_{57}$}
    \uput[90](2.2,3.8){$\boldsymbol{\beta}_-^+$}
    \uput[90](7.2,3.8){$\boldsymbol{\beta}_{56}$}
\end{pspicture}
\end{center}
\caption{The complete set of screenings for the first deformation of the $OSp(7|2)$ sigma model.}
\label{OSp(7|2)_screening_set_first_deformation}
\end{figure}

\begin{figure}[H]
\begin{center}
\begin{pspicture}(0,0.5)(9.5,4.5)
    \pscircle[linewidth=1.5pt](1,1){0.2}
    \psline[linewidth=1.5pt](0.87,0.87)(1.13,1.13)
    \psline[linewidth=1.5pt](0.87,1.13)(1.13,0.87)
    \pscircle[linewidth=1.5pt](8.5,1){0.2}
    \psline[linewidth=1.5pt](8.37,0.87)(8.63,1.13)
    \psline[linewidth=1.5pt](8.37,1.13)(8.63,0.87)
    \pscircle[linewidth=1.5pt](1,4){0.2}
    \psline[linewidth=1.5pt](0.87,3.87)(1.13,4.13)
    \psline[linewidth=1.5pt](0.87,4.13)(1.13,3.87)
    \pscircle[linewidth=1.5pt](8.5,4){0.2}
    \psline[linewidth=1.5pt](8.37,3.87)(8.63,4.13)
    \psline[linewidth=1.5pt](8.37,4.13)(8.63,3.87)
    \pscircle[linewidth=1.5pt](2.5,2.5){0.2}
    \psline[linewidth=1.5pt](2.37,2.37)(2.63,2.63)
    \psline[linewidth=1.5pt](2.37,2.63)(2.63,2.37)
    \pscircle[linewidth=1.5pt](4,2.5){0.2}
    \psline[linewidth=1.5pt](3.87,2.37)(4.13,2.63)
    \psline[linewidth=1.5pt](3.87,2.63)(4.13,2.37)
    \pscircle[linewidth=1.5pt](5.5,2.5){0.2}
    \psline[linewidth=1.5pt](5.37,2.37)(5.63,2.63)
    \psline[linewidth=1.5pt](5.37,2.63)(5.63,2.37)
    \pscircle[linewidth=1.5pt](7,2.5){0.2}
    \psline[linewidth=1.5pt](6.87,2.37)(7.13,2.63)
    \psline[linewidth=1.5pt](6.87,2.63)(7.13,2.37)
    \pscircle[linewidth=1.5pt](3.25,3.3){0.2}
    \psline[linewidth=1.5pt](3.15,3.6)(3.35,3.6)
    \pscircle[linewidth=1.5pt](0.2,2.5){0.2}
    \psline[linewidth=1.5pt](0.1,2.8)(0.3,2.8)
    \pscircle[linewidth=1.5pt](9.3,2.5){0.2}
    \psline[linewidth=1.5pt](9.2,2.8)(9.4,2.8)
    \pscircle[linewidth=1.5pt](2.2,3.6){0.2}
    \psline[linewidth=1.5pt](2.1,3.9)(2.3,3.9)
    \pscircle[linewidth=1.5pt](2.2,1.4){0.2}
    \psline[linewidth=1.5pt](2.1,1.7)(2.3,1.7)
    \pscircle[linewidth=1.5pt](7.2,3.6){0.2}
    \psline[linewidth=1.5pt](7.07,3.47)(7.33,3.73)
    \psline[linewidth=1.5pt](7.07,3.73)(7.33,3.47)
    \psline[linewidth=1.5pt](7.1,3.9)(7.3,3.9)
    \pscircle[linewidth=1.5pt](7.2,1.4){0.2}
    \psline[linewidth=1.5pt](7.07,1.27)(7.33,1.53)
    \psline[linewidth=1.5pt](7.07,1.53)(7.33,1.27)
    \psline[linewidth=1.5pt](7.1,1.7)(7.3,1.7)
    \pscircle[linewidth=1.5pt](4.75,3.3){0.2}
    \psline[linewidth=1.5pt](4.62,3.17)(4.88,3.43)
    \psline[linewidth=1.5pt](4.62,3.43)(4.88,3.17)
    \psline[linewidth=1.5pt](4.65,3.6)(4.85,3.6)
    \pscircle[linewidth=1.5pt](6.25,3.3){0.2}
    \psline[linewidth=1.5pt](6.12,3.17)(6.38,3.43)
    \psline[linewidth=1.5pt](6.12,3.43)(6.38,3.17)
    \psline[linewidth=1.5pt](6.15,3.6)(6.35,3.6)
    \psline[linewidth=1.5pt](1,1.2)(1,3.8)
    \psline[linewidth=1.5pt](8.5,1.2)(8.5,3.8)
    \psline[linewidth=1.5pt](1.13,1.13)(2.37,2.37)
    \psline[linewidth=1.5pt](1.13,3.87)(2.37,2.63)
    \psline[linewidth=1.5pt,linestyle=dashed](7.13,2.63)(8.37,3.87)
    \psline[linewidth=1.5pt,linestyle=dashed](7.13,2.37)(8.37,1.13)
    \psline[linewidth=1.5pt](2.7,2.5)(3.8,2.5)
    \psline[linewidth=1.5pt,linestyle=dashed](4.2,2.5)(5.3,2.5)
    \psline[linewidth=1.5pt,linestyle=dashed](5.7,2.5)(6.8,2.5)
    \uput[-90](1,0.8){$\boldsymbol{\alpha}_1$}
    \uput[90](1,4.2){$\boldsymbol{\alpha}_2$}
    \uput[-90](2.5,2.3){$\boldsymbol{\alpha}_3$}
    \uput[90](3.25,3.5){$\boldsymbol{\beta}_{34}$}
    \uput[-90](4,2.3){$\boldsymbol{\alpha}_4$}
    \uput[90](4.75,3.5){$\boldsymbol{\beta}_-$}
    \uput[-90](5.5,2.3){$\boldsymbol{\beta}_1$}
    \uput[90](6.25,3.5){$\boldsymbol{\alpha}_{5}$}
    \uput[-90](7,2.3){$\boldsymbol{\beta}_2$}
    \uput[90](8.5,4.2){$\boldsymbol{\alpha}_6$}
    \uput[-90](8.5,0.8){$\boldsymbol{\alpha}_7$}
    \uput[-90](0.2,2.3){$\boldsymbol{\beta}_{12}$}
    \uput[-90](9.3,2.3){$\boldsymbol{\beta}_{67}$}
    \uput[-90](2.2,1.3){$\boldsymbol{\beta}_{13}$}
    \uput[-90](7.2,1.3){$\boldsymbol{\beta}_+^-$}
    \uput[90](2.2,3.8){$\boldsymbol{\beta}_{23}$}
    \uput[90](7.2,3.8){$\boldsymbol{\beta}_+^+$}
\end{pspicture}
\end{center}
\caption{The complete set of screenings for the second deformation of the $OSp(7|2)$ sigma model.}
\label{OSp(7|2)_screening_set_second_deformation}
\end{figure}
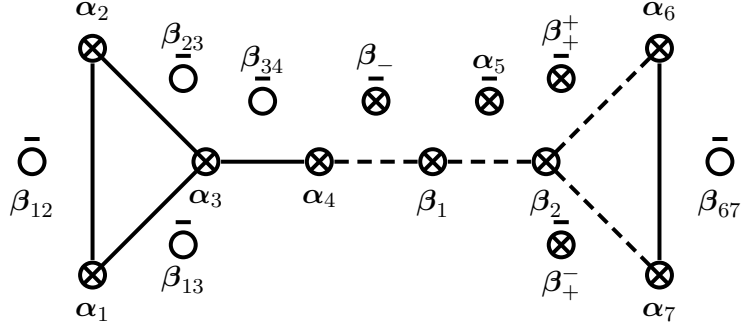

Let us now turn to the deformed $OSp(7|2)$ sigma model. There exist at least two integrable deformations of the $OSp(7|2)$ sigma model, the first of which is described by Fig. \ref{OSp(7|2)_screening_set_first_deformation},
and the second one is described by the screenings in Fig. \ref{OSp(7|2)_screening_set_second_deformation}.

In a similar manner, the metric and the $b \rightarrow 0$ limit for the $OSp(7|2)$ sigma model can be described. The metric of both deformations of the $OSp(7|2)$ sigma model is specified by the screenings in Fig. \ref{OSp(7|2)_sigma_model_screening_set_first_deformation} and \ref{OSp(7|2)_sigma_model_screening_set_second_deformation}, respectively.
\begin{figure}[H]
\begin{minipage}{0.45\linewidth}
\centering
\begin{pspicture}(0,0.5)(7,4.5)
    \pscircle[linewidth=1.5pt](1,1){0.2}
    \psline[linewidth=1.5pt](0.9,1.3)(1.1,1.3)
    \pscircle[linewidth=1.5pt](1,4){0.2}
    \psline[linewidth=1.5pt](0.9,4.3)(1.1,4.3)
    \pscircle[linewidth=1.5pt](4,2.5){0.2}
    \psline[linewidth=1.5pt](3.87,2.37)(4.13,2.63)
    \psline[linewidth=1.5pt](3.87,2.63)(4.13,2.37)
    \pscircle[linewidth=1.5pt](5.5,2.5){0.2}
    \psline[linewidth=1.5pt](5.4,2.8)(5.6,2.8)
    \pscircle[linewidth=1.5pt](2.5,2.5){0.2}
    \psline[linewidth=1.5pt](2.4,2.8)(2.6,2.8)
    \psline[linewidth=1.5pt](1.13,1.13)(2.63,2.63)
    \psline[linewidth=1.5pt](1.13,3.87)(2.63,2.37)
    \psline[linewidth=1.5pt](2.63,2.5)(3.87,2.5)
    \psline[linewidth=1.5pt](4.13,2.5)(5.37,2.5)
    \uput[180](0.7,1){$\boldsymbol{\beta}_{56}$}
    \uput[180](0.7,4){$\boldsymbol{\beta}_{57}$}
    \uput[-90](2.5,2.3){$\boldsymbol{\beta}_+$}
    \uput[-90](4,2.3){$\boldsymbol{\beta}_1$}
    \uput[-90](5.5,2.3){$\boldsymbol{\beta}_{12}$}
    \uput[90](3.25,2.5){$1+\frac{1}{b^2}$}
    \uput[90](4.75,2.5){$-\frac{1}{b^2}$}
    \uput[45](1.6,3.2){$-\frac{1}{b^2}$}
    \uput[-45](1.6,1.8){$-\frac{1}{b^2}$}
\end{pspicture}
\caption{The screening set in the ultraviolet limit for the first deformation of the $OSp(7|2)$ sigma model.}
\label{OSp(7|2)_sigma_model_screening_set_first_deformation}
\end{minipage}
\hfill
\begin{minipage}{0.45\linewidth}
\centering
\begin{pspicture}(0,0.5)(7,4.5)
    \pscircle[linewidth=1.5pt](1,1){0.2}
    \psline[linewidth=1.5pt](0.87,0.87)(1.13,1.13)
    \psline[linewidth=1.5pt](0.87,1.13)(1.13,0.87)
    \psline[linewidth=1.5pt](0.9,1.3)(1.1,1.3)
    \pscircle[linewidth=1.5pt](1,4){0.2}
    \psline[linewidth=1.5pt](0.87,3.87)(1.13,4.13)
    \psline[linewidth=1.5pt](0.87,4.13)(1.13,3.87)
    \psline[linewidth=1.5pt](0.9,4.3)(1.1,4.3)
    \pscircle[linewidth=1.5pt](4,2.5){0.2}
    \psline[linewidth=1.5pt](3.9,2.8)(4.1,2.8)
    \pscircle[linewidth=1.5pt](2.5,2.5){0.2}
    \pscircle[linewidth=1.5pt](5.5,2.5){0.2}
    \psline[linewidth=1.5pt](5.4,2.8)(5.6,2.8)
    \psline[linewidth=1.5pt](1.13,1.13)(2.63,2.63)
    \psline[linewidth=1.5pt](1.13,3.87)(2.63,2.37)
    \psline[linewidth=1.5pt](2.63,2.5)(3.87,2.5)
    \psline[linewidth=1.5pt](4.13,2.5)(5.37,2.5)
    \uput[180](0.7,1){$\boldsymbol{\beta}_+^-$}
    \uput[180](0.7,4){$\boldsymbol{\beta}_+^+$}
    \uput[-90](2.5,2.3){$\boldsymbol{\beta}_1$}
    \uput[-90](4,2.3){$\boldsymbol{\beta}_{34}$}
    \uput[-90](5.5,2.3){$\boldsymbol{\beta}_{12}$}
    \uput[90](3.25,2.5){$-\frac{1}{b^2}$}
    \uput[90](4.75,2.5){$-\frac{1}{b^2}$}
    \uput[45](1.6,3.2){$1+\frac{1}{b^2}$}
    \uput[-45](1.6,1.8){$1+\frac{1}{b^2}$}
\end{pspicture}
\caption{The screening set in the ultraviolet limit for the second deformation of the $OSp(7|2)$ sigma model.}
\label{OSp(7|2)_sigma_model_screening_set_second_deformation}
\end{minipage}
\end{figure}

And in the limit $b \rightarrow 0$, we obtain the following screening charges in Fig. \ref{OSp(7|2)_Toda_screening_set_first_deformation} and \ref{OSp(7|2)_sigma_model_screening_set_second_deformation}, respectively,
\begin{figure}[H]
\begin{minipage}{0.45\linewidth}
\centering
\begin{pspicture}(0.5,0.5)(7.5,4.5)
    \pscircle[linewidth=1.5pt](1,1){0.2}
    \psline[linewidth=1.5pt](0.87,0.87)(1.13,1.13)
    \psline[linewidth=1.5pt](0.87,1.13)(1.13,0.87)
    \pscircle[linewidth=1.5pt](1,4){0.2}
    \psline[linewidth=1.5pt](0.87,3.87)(1.13,4.13)
    \psline[linewidth=1.5pt](0.87,4.13)(1.13,3.87)
    \pscircle[linewidth=1.5pt](2.5,2.5){0.2}
    \psline[linewidth=1.5pt](2.4,2.8)(2.6,2.8)
    \pscircle[linewidth=1.5pt](4,2.5){0.2}
    \psline[linewidth=1.5pt](3.87,2.37)(4.13,2.63)
    \psline[linewidth=1.5pt](3.87,2.63)(4.13,2.37)
    \pscircle[linewidth=1.5pt](5.5,2.5){0.2}
    \psline[linewidth=1.5pt](5.37,2.37)(5.63,2.63)
    \psline[linewidth=1.5pt](5.37,2.63)(5.63,2.37)
    \pscircle[linewidth=1.5pt](7,1){0.2}
    \psline[linewidth=1.5pt](6.87,0.87)(7.13,1.13)
    \psline[linewidth=1.5pt](6.87,1.13)(7.13,0.87)
    \pscircle[linewidth=1.5pt](7,4){0.2}
	\psline[linewidth=1.5pt](6.87,3.87)(7.13,4.13)    
    \psline[linewidth=1.5pt](6.87,4.13)(7.13,3.87)
    \psline[linewidth=1.5pt](1.13,1.13)(2.63,2.63)
    \psline[linewidth=1.5pt](1.13,3.87)(2.63,2.37)
    \psline[linewidth=1.5pt](2.63,2.5)(3.87,2.5)
    \psline[linewidth=1.5pt](4.13,2.5)(5.37,2.5)
    \psline[linewidth=1.5pt](5.63,2.63)(6.87,3.87)
    \psline[linewidth=1.5pt](5.63,2.37)(6.87,1.13)
    \psline[linewidth=1.5pt](1,1.2)(1,3.8)
    \psline[linewidth=1.5pt](7,1.2)(7,3.8)
    \uput[-90](1,0.8){$\boldsymbol{\alpha}_1$}
    \uput[90](1,4.1){$\boldsymbol{\alpha}_2$}
    \uput[-90](2.5,2.3){$\boldsymbol{\alpha}_3$}
    \uput[-90](4,2.3){$\boldsymbol{\alpha}_4$}
    \uput[-90](5.5,2.3){$\boldsymbol{\alpha}_5$}
    \uput[90](7,4.1){$\boldsymbol{\alpha}_6$}
    \uput[-90](7,0.8){$\boldsymbol{\alpha}_7$}
\end{pspicture}
\caption{The screening set for the dual description of the first deformation of the $OSp(7|2)$ sigma model in the limit $b \rightarrow 0$.}
\label{OSp(7|2)_Toda_screening_set_first_deformation}
\end{minipage}
\hfill
\begin{minipage}{0.45\linewidth}
\begin{pspicture}(0.5,0.5)(7.5,4.5)
    \pscircle[linewidth=1.5pt](1,1){0.2}
    \psline[linewidth=1.5pt](0.87,0.87)(1.13,1.13)
    \psline[linewidth=1.5pt](0.87,1.13)(1.13,0.87)
    \pscircle[linewidth=1.5pt](1,4){0.2}
    \psline[linewidth=1.5pt](0.87,3.87)(1.13,4.13)
    \psline[linewidth=1.5pt](0.87,4.13)(1.13,3.87)
    \pscircle[linewidth=1.5pt](2.5,2.5){0.2}
    \psline[linewidth=1.5pt](5.4,2.8)(5.6,2.8)
    \pscircle[linewidth=1.5pt](4,2.5){0.2}
    \psline[linewidth=1.5pt](3.87,2.37)(4.13,2.63)
    \psline[linewidth=1.5pt](3.87,2.63)(4.13,2.37)
    \pscircle[linewidth=1.5pt](5.5,2.5){0.2}
    \psline[linewidth=1.5pt](5.37,2.37)(5.63,2.63)
    \psline[linewidth=1.5pt](5.37,2.63)(5.63,2.37)
    \pscircle[linewidth=1.5pt](7,1){0.2}
    \psline[linewidth=1.5pt](6.87,0.87)(7.13,1.13)
    \psline[linewidth=1.5pt](6.87,1.13)(7.13,0.87)
    \pscircle[linewidth=1.5pt](7,4){0.2}
	\psline[linewidth=1.5pt](6.87,3.87)(7.13,4.13)    
    \psline[linewidth=1.5pt](6.87,4.13)(7.13,3.87)
    \psline[linewidth=1.5pt](1.13,1.13)(2.63,2.63)
    \psline[linewidth=1.5pt](1.13,3.87)(2.63,2.37)
    \psline[linewidth=1.5pt](2.63,2.5)(3.87,2.5)
    \psline[linewidth=1.5pt](4.13,2.5)(5.37,2.5)
    \psline[linewidth=1.5pt](5.63,2.63)(6.87,3.87)
    \psline[linewidth=1.5pt](5.63,2.37)(6.87,1.13)
    \psline[linewidth=1.5pt](1,1.2)(1,3.8)
    \psline[linewidth=1.5pt](7,1.2)(7,3.8)
    \uput[-90](1,0.8){$\boldsymbol{\alpha}_1$}
    \uput[90](1,4.1){$\boldsymbol{\alpha}_2$}
    \uput[-90](2.5,2.3){$\boldsymbol{\alpha}_3$}
    \uput[-90](4,2.3){$\boldsymbol{\alpha}_4$}
    \uput[-90](5.5,2.3){$\boldsymbol{\alpha}_5$}
    \uput[90](7,4.1){$\boldsymbol{\alpha}_6$}
    \uput[-90](7,0.8){$\boldsymbol{\alpha}_7$}
\end{pspicture}
\caption{The screening set for the dual description of the second deformation of the $OSp(7|2)$ sigma model in the limit $b \rightarrow 0$.}
\label{OSp(7|2)_Toda_screening_set_second_deformation}
\end{minipage}
\end{figure}
\noindent which are the screening systems for the deformed $O(7)$ sigma model that are regular in the limit $b \rightarrow 0$, with the only difference being that the screenings $\boldsymbol{\alpha}_3$ and $\boldsymbol{\alpha}_5$, respectively, become dressed.

\begin{figure}[H]
\begin{center}
\begin{pspicture}(0.5,0.5)(9,4.5)
    \pscircle[linewidth=1.5pt](1,1){0.2}
    \psline[linewidth=1.5pt](0.87,0.87)(1.13,1.13)
    \psline[linewidth=1.5pt](0.87,1.13)(1.13,0.87)
    \pscircle[linewidth=1.5pt](1,4){0.2}
    \psline[linewidth=1.5pt](0.87,3.87)(1.13,4.13)
    \psline[linewidth=1.5pt](0.87,4.13)(1.13,3.87)
    \pscircle[linewidth=1.5pt](2.5,2.5){0.2}
    \psline[linewidth=1.5pt](2.37,2.37)(2.63,2.63)
    \psline[linewidth=1.5pt](2.37,2.63)(2.63,2.37)
    \pscircle[linewidth=1.5pt](4,2.5){0.2}
    \psline[linewidth=1.5pt](3.87,2.37)(4.13,2.63)
    \psline[linewidth=1.5pt](3.87,2.63)(4.13,2.37)
    \pscircle[linewidth=1.5pt](5.5,2.5){0.2}
    \psline[linewidth=1.5pt](5.37,2.37)(5.63,2.63)
    \psline[linewidth=1.5pt](5.37,2.63)(5.63,2.37)
    \pscircle[linewidth=1.5pt](7,2.5){0.2}
    \psline[linewidth=1.5pt](6.87,2.37)(7.13,2.63)
    \psline[linewidth=1.5pt](6.87,2.63)(7.13,2.37)
    \pscircle[linewidth=1.5pt](8.5,2.5){0.2}
    \psline[linewidth=1.5pt](8.37,2.37)(8.63,2.63)
    \psline[linewidth=1.5pt](8.37,2.63)(8.63,2.37)
    \psline[linewidth=1.5pt](1,1.2)(1,3.8)
    \psline[linewidth=1.5pt](1.13,1.13)(2.37,2.37)
    \psline[linewidth=1.5pt](1.13,3.87)(2.37,2.63)
    \psline[linewidth=1.5pt,linestyle=dashed](2.7,2.5)(3.8,2.5)
    \psline[linewidth=1.5pt,linestyle=dashed](4.2,2.5)(5.3,2.5)
    \psline[linewidth=1.5pt,linestyle=dashed](5.7,2.5)(6.8,2.5)
    \psline[linewidth=1.5pt](7.2,2.5)(8.3,2.5)
    \uput[-90](1,0.8){$\boldsymbol{\alpha}_1$}
    \uput[90](1,4.2){$\boldsymbol{\alpha}_2$}
    \uput[-90](2.5,2.3){$\boldsymbol{\alpha}_3$}
    \uput[-90](4,2.3){$-\boldsymbol{\beta}_2$}
    \uput[-90](5.5,2.3){$\boldsymbol{\beta}_3$}
    \uput[-90](7,2.3){$\boldsymbol{\alpha}_5$}
    \uput[-90](8.5,2.3){$\boldsymbol{\alpha}_6$}
\end{pspicture}
\end{center}
\caption{The screening set for the third realization of the $\frac{\widehat{\mathfrak{osp}}(7|2)_w}{\widehat{\mathfrak{osp}}(7|2)_w}$ coset conformal field theory.}
\label{osp(7|2)_cft_screening_set_third_realization}
\end{figure}
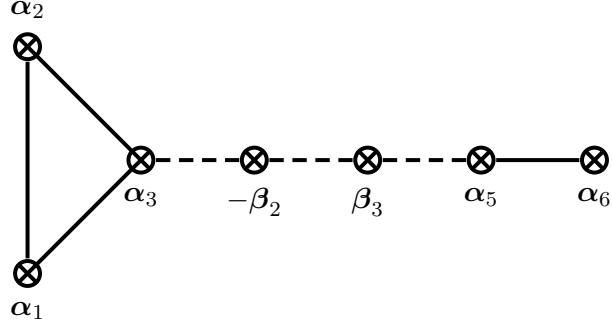

Similarly to the previous case of $OSp(5|2)$, we can write the screening systems for the $\frac{\widehat{\mathfrak{osp}}(7|2)_w}{\widehat{\mathfrak{osp}}(6|2)_w}$ coset conformal field theory, which look as follows in Fig. \ref{osp(7|2)_cft_screening_set_first_realization} and \ref{osp(7|2)_cft_screening_set_second_realization}.
\begin{figure}[H]
\begin{minipage}{0.45\linewidth}
\centering
\begin{pspicture}(0.5,0.5)(9,4.5)
    \pscircle[linewidth=1.5pt](1,1){0.2}
    \psline[linewidth=1.5pt](0.87,0.87)(1.13,1.13)
    \psline[linewidth=1.5pt](0.87,1.13)(1.13,0.87)
    \pscircle[linewidth=1.5pt](1,4){0.2}
    \psline[linewidth=1.5pt](0.87,3.87)(1.13,4.13)
    \psline[linewidth=1.5pt](0.87,4.13)(1.13,3.87)
    \pscircle[linewidth=1.5pt](2.5,2.5){0.2}
    \psline[linewidth=1.5pt](2.37,2.37)(2.63,2.63)
    \psline[linewidth=1.5pt](2.37,2.63)(2.63,2.37)
    \pscircle[linewidth=1.5pt](4,2.5){0.2}
    \psline[linewidth=1.5pt](3.87,2.37)(4.13,2.63)
    \psline[linewidth=1.5pt](3.87,2.63)(4.13,2.37)
    \pscircle[linewidth=1.5pt](5.5,2.5){0.2}
    \psline[linewidth=1.5pt](5.37,2.37)(5.63,2.63)
    \psline[linewidth=1.5pt](5.37,2.63)(5.63,2.37)
    \pscircle[linewidth=1.5pt](7,2.5){0.2}
    \psline[linewidth=1.5pt](6.87,2.37)(7.13,2.63)
    \psline[linewidth=1.5pt](6.87,2.63)(7.13,2.37)
    \pscircle[linewidth=1.5pt](8.5,2.5){0.2}
    \psline[linewidth=1.5pt](8.37,2.37)(8.63,2.63)
    \psline[linewidth=1.5pt](8.37,2.63)(8.63,2.37)
    \psline[linewidth=1.5pt](1,1.2)(1,3.8)
    \psline[linewidth=1.5pt,linestyle=dashed](1.13,1.13)(2.37,2.37)
    \psline[linewidth=1.5pt,linestyle=dashed](1.13,3.87)(2.37,2.63)
    \psline[linewidth=1.5pt,linestyle=dashed](2.7,2.5)(3.8,2.5)
    \psline[linewidth=1.5pt,linestyle=dashed](4.2,2.5)(5.3,2.5)
    \psline[linewidth=1.5pt](5.7,2.5)(6.8,2.5)
    \psline[linewidth=1.5pt](7.2,2.5)(8.3,2.5)
    \uput[-90](1,0.8){$\boldsymbol{\alpha}_1$}
    \uput[90](1,4.2){$\boldsymbol{\alpha}_2$}
    \uput[-90](2.5,2.3){$\boldsymbol{\beta}_1$}
    \uput[-90](4,2.3){$\boldsymbol{\beta}_2$}
    \uput[-90](5.5,2.3){$\boldsymbol{\alpha}_4$}
    \uput[-90](7,2.3){$\boldsymbol{\alpha}_5$}
    \uput[-90](8.5,2.3){$\boldsymbol{\alpha}_6$}
\end{pspicture}
\caption{The screening set for the first realization of the $\frac{\widehat{\mathfrak{osp}}(7|2)_w}{\widehat{\mathfrak{osp}}(7|2)_w}$ coset conformal field theory.}
\label{osp(7|2)_cft_screening_set_first_realization}
\end{minipage}
\hfill
\begin{minipage}{0.45\linewidth}
\centering
\begin{pspicture}(0.5,0.5)(9,4.5)
    \pscircle[linewidth=1.5pt](1,1){0.2}
    \psline[linewidth=1.5pt](0.87,0.87)(1.13,1.13)
    \psline[linewidth=1.5pt](0.87,1.13)(1.13,0.87)
    \pscircle[linewidth=1.5pt](1,4){0.2}
    \psline[linewidth=1.5pt](0.87,3.87)(1.13,4.13)
    \psline[linewidth=1.5pt](0.87,4.13)(1.13,3.87)
    \pscircle[linewidth=1.5pt](2.5,2.5){0.2}
    \psline[linewidth=1.5pt](2.37,2.37)(2.63,2.63)
    \psline[linewidth=1.5pt](2.37,2.63)(2.63,2.37)
    \pscircle[linewidth=1.5pt](4,2.5){0.2}
    \psline[linewidth=1.5pt](3.87,2.37)(4.13,2.63)
    \psline[linewidth=1.5pt](3.87,2.63)(4.13,2.37)
    \pscircle[linewidth=1.5pt](5.5,2.5){0.2}
    \psline[linewidth=1.5pt](5.37,2.37)(5.63,2.63)
    \psline[linewidth=1.5pt](5.37,2.63)(5.63,2.37)
    \pscircle[linewidth=1.5pt](7,2.5){0.2}
    \psline[linewidth=1.5pt](6.87,2.37)(7.13,2.63)
    \psline[linewidth=1.5pt](6.87,2.63)(7.13,2.37)
    \pscircle[linewidth=1.5pt](8.5,2.5){0.2}
    \psline[linewidth=1.5pt](8.37,2.37)(8.63,2.63)
    \psline[linewidth=1.5pt](8.37,2.63)(8.63,2.37)
    \psline[linewidth=1.5pt](1,1.2)(1,3.8)
    \psline[linewidth=1.5pt](1.13,1.13)(2.37,2.37)
    \psline[linewidth=1.5pt](1.13,3.87)(2.37,2.63)
    \psline[linewidth=1.5pt](2.7,2.5)(3.8,2.5)
    \psline[linewidth=1.5pt,linestyle=dashed](4.2,2.5)(5.3,2.5)
    \psline[linewidth=1.5pt,linestyle=dashed](5.7,2.5)(6.8,2.5)
    \psline[linewidth=1.5pt,linestyle=dashed](7.2,2.5)(8.3,2.5)
    \uput[-90](1,0.8){$\boldsymbol{\alpha}_1$}
    \uput[90](1,4.2){$\boldsymbol{\alpha}_2$}
    \uput[-90](2.5,2.3){$\boldsymbol{\alpha}_3$}
    \uput[-90](4,2.3){$\boldsymbol{\alpha}_4$}
    \uput[-90](5.5,2.3){$\boldsymbol{\beta}_1$}
    \uput[-90](7,2.3){$\boldsymbol{\beta}_2$}
    \uput[-90](8.5,2.3){$\boldsymbol{\alpha}_6$}
\end{pspicture}
\caption{The screening set for the second realization of the $\frac{\widehat{\mathfrak{osp}}(7|2)_w}{\widehat{\mathfrak{osp}}(7|2)_w}$ coset conformal field theory.
}
\label{osp(7|2)_cft_screening_set_second_realization}
\end{minipage}
\end{figure}
In light of fermionic reflections, an important point can be noted about the previous two diagrams: if one first performs a reflection with respect to the root $\boldsymbol{\beta}_2$, the diagram in Fig. \ref{osp(7|2)_cft_screening_set_first_realization} turns into the diagram in Fig. \ref{osp(7|2)_cft_screening_set_third_realization},
\noindent which, after a reflection with respect to the root $\boldsymbol{\beta}_3$, turns into the diagram in Fig. \ref{osp(7|2)_cft_screening_set_second_realization}. Exactly the same statement holds for affine diagrams with the root $\boldsymbol{\alpha}_7$; however, for such integrable but, generally speaking, non-conformal theories, the possibility of fermionic reflections has not been explicitly verified, although it is beyond doubt. In simple terms, fermionic reflections shift the connections between the screenings, depicted by dashed lines, to the right.

\begin{figure}[H]
\begin{center}
\begin{pspicture}(0.5,0.5)(7.5,4.5)
	\pscircle[linewidth=1.5pt](1,1){0.2}
    \pscircle[linewidth=1.5pt](2.5,1){0.2}
    \pscircle[linewidth=1.5pt](4,1){0.2}
    \psline[linewidth=1.5pt](3.87,0.87)(4.13,1.13)
    \psline[linewidth=1.5pt](3.87,1.13)(4.13,0.87)
    \pscircle[linewidth=1.5pt](5.5,2.5){0.2}
    \psline[linewidth=1.5pt](5.37,2.37)(5.63,2.63)
    \psline[linewidth=1.5pt](5.37,2.63)(5.63,2.37)
    \pscircle[linewidth=1.5pt](7,1){0.2}
    \pscircle[linewidth=1.5pt](7,4){0.2}
	\psline[linewidth=1.5pt,doubleline=true,doublesep=1pt]{<-}(1.2,1)(2.3,1)
	\psline[linewidth=1.5pt](2.7,1)(3.8,1)
	\psline[linewidth=1.5pt,linestyle=dashed](4.13,1.13)(5.37,2.37)
	\psline[linewidth=1.5pt](5.63,2.63)(6.87,3.87)
	\psline[linewidth=1.5pt](5.63,2.37)(6.87,1.13)
	\psframe[linewidth=1.5pt,linestyle=dashed](0.6,0.4)(7.4,2.9)
	\psframe[linewidth=1.5pt,linestyle=dotted](2.1,0.6)(7.6,4.4)
	\uput[90](1.75,1.1){$\frac{2\beta^2}{b^2}$}
	\uput[90](3.25,1.1){$\frac{\beta^2}{b^2}$}
	\uput[135](6.2,3.3){$\beta^2$}
	\uput[-135](6.2,1.7){$\beta^2$}
\end{pspicture}
\caption{The general diagram featuring different realizations of the $\frac{\widehat{\mathfrak{osp}}(5|2)_w}{\widehat{\mathfrak{osp}}(4|2)_w}$ coset conformal field theory.}
\label{osp(5|2)_cft_both_realizations}
\end{center}
\end{figure}
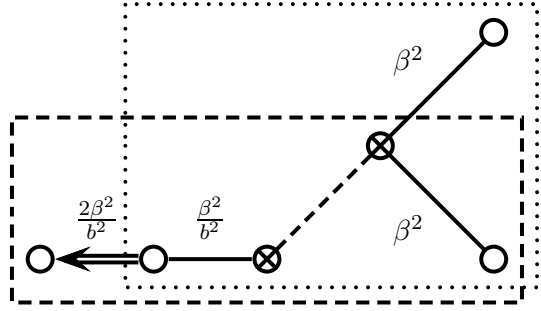

Now we are ready to formulate our hypothesis for the set of screening charges of an arbitrary deformed $OSp(N|2m)$ sigma model. Let us first focus on the $OSp(N|2)$ case. For the $OSp(5|2)$ case, let us draw the diagram in Fig. \ref{osp(5|2)_cft_both_realizations} as follows: we place the diagram corresponding in shape to the Dynkin diagram for the Lie superalgebra $\mathfrak{osp}(5|2)$ at the top, below it we place the diagram in the shape of the Dynkin diagram for $\mathfrak{osp}(4|2)$, and we connect them with a dashed line
\noindent then, by applying fermionic reflections to the diagrams inside the dashed and dotted boxes respectively, we obtain the diagram describing the $\frac{\widehat{\mathfrak{osp}}(5|2)_w}{\widehat{\mathfrak{osp}}(4|2)_w}$ coset conformal field theory in Fig. \ref{osp(5|2)_cft_screeinig_set}.

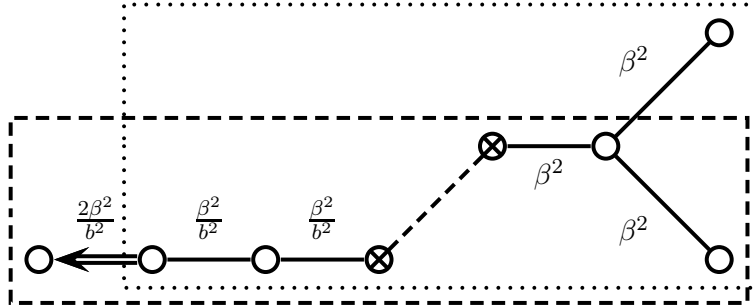
\begin{figure}[H]
\begin{center}
\begin{pspicture}(0.5,0.5)(10.5,4.5)
	\pscircle[linewidth=1.5pt](1,1){0.2}
    \pscircle[linewidth=1.5pt](2.5,1){0.2}
    \pscircle[linewidth=1.5pt](4,1){0.2}
    \pscircle[linewidth=1.5pt](5.5,1){0.2}
    \psline[linewidth=1.5pt](5.37,0.87)(5.63,1.13)
    \psline[linewidth=1.5pt](5.37,1.13)(5.63,0.87)
    \pscircle[linewidth=1.5pt](7,2.5){0.2}
    \psline[linewidth=1.5pt](6.87,2.37)(7.13,2.63)
    \psline[linewidth=1.5pt](6.87,2.63)(7.13,2.37)
    \pscircle[linewidth=1.5pt](8.5,2.5){0.2}
    \pscircle[linewidth=1.5pt](10,1){0.2}
    \pscircle[linewidth=1.5pt](10,4){0.2}
	\psline[linewidth=1.5pt,doubleline=true,doublesep=1pt]{<-}(1.2,1)(2.3,1)
	\psline[linewidth=1.5pt](2.7,1)(3.8,1)
	\psline[linewidth=1.5pt](4.2,1)(5.3,1)
	\psline[linewidth=1.5pt,linestyle=dashed](5.63,1.13)(6.87,2.37)
	\psline[linewidth=1.5pt](7.2,2.5)(8.3,2.5)
	\psline[linewidth=1.5pt](8.63,2.63)(9.87,3.87)
	\psline[linewidth=1.5pt](8.63,2.37)(9.87,1.13)
	\psframe[linewidth=1.5pt,linestyle=dashed](0.6,0.4)(10.4,2.9)
	\psframe[linewidth=1.5pt,linestyle=dotted](2.1,0.6)(10.6,4.4)
	\uput[90](1.75,1.1){$\frac{2\beta^2}{b^2}$}
	\uput[90](3.25,1.1){$\frac{\beta^2}{b^2}$}
	\uput[90](4.75,1.1){$\frac{\beta^2}{b^2}$}
	\uput[-90](7.75,2.5){$\beta^2$}
	\uput[135](9.2,3.3){$\beta^2$}
	\uput[-135](9.2,1.7){$\beta^2$}
\end{pspicture}
\end{center}
\caption{The general diagram featuring different realizations of the $\frac{\widehat{\mathfrak{osp}}(7|2)_w}{\widehat{\mathfrak{osp}}(6|2)_w}$ coset conformal field theory.}
\label{osp(7|2)_cft_screening_set_both_realizations}
\end{figure}

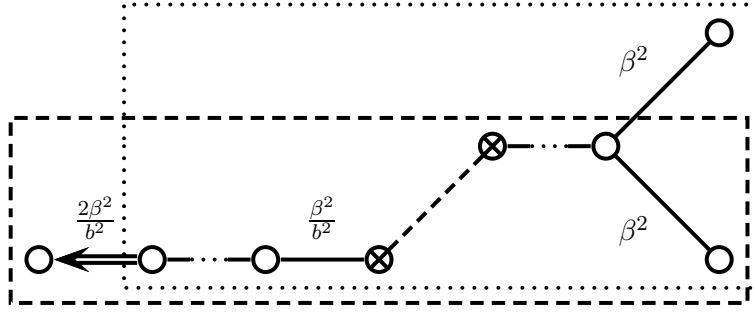
\begin{figure}[H]
\begin{center}
\begin{pspicture}(0.5,0.5)(10.5,4.5)
	\pscircle[linewidth=1.5pt](1,1){0.2}
    \pscircle[linewidth=1.5pt](2.5,1){0.2}
    \pscircle[linewidth=1.5pt](4,1){0.2}
    \pscircle[linewidth=1.5pt](5.5,1){0.2}
    \psline[linewidth=1.5pt](5.37,0.87)(5.63,1.13)
    \psline[linewidth=1.5pt](5.37,1.13)(5.63,0.87)
    \pscircle[linewidth=1.5pt](7,2.5){0.2}
    \psline[linewidth=1.5pt](6.87,2.37)(7.13,2.63)
    \psline[linewidth=1.5pt](6.87,2.63)(7.13,2.37)
    \pscircle[linewidth=1.5pt](8.5,2.5){0.2}
    \pscircle[linewidth=1.5pt](10,1){0.2}
    \pscircle[linewidth=1.5pt](10,4){0.2}
	\psline[linewidth=1.5pt,doubleline=true,doublesep=1pt]{<-}(1.2,1)(2.3,1)
	\psline[linewidth=1.5pt](2.7,1)(3.0,1)
	\psline[linewidth=1.5pt,linestyle=dotted](3.0,1)(3.5,1)
	\psline[linewidth=1.5pt](3.5,1)(3.8,1)
	\psline[linewidth=1.5pt](4.2,1)(5.3,1)
	\psline[linewidth=1.5pt,linestyle=dashed](5.63,1.13)(6.87,2.37)
	\psline[linewidth=1.5pt](7.2,2.5)(7.5,2.5)
	\psline[linewidth=1.5pt,linestyle=dotted](7.5,2.5)(8.0,2.5)
	\psline[linewidth=1.5pt](8.0,2.5)(8.3,2.5)
	\psline[linewidth=1.5pt](8.63,2.63)(9.87,3.87)
	\psline[linewidth=1.5pt](8.63,2.37)(9.87,1.13)
	\psframe[linewidth=1.5pt,linestyle=dashed](0.6,0.4)(10.4,2.9)
	\psframe[linewidth=1.5pt,linestyle=dotted](2.1,0.6)(10.6,4.4)
	\uput[90](1.75,1.1){$\frac{2\beta^2}{b^2}$}
	\uput[90](4.75,1.1){$\frac{\beta^2}{b^2}$}
	\uput[135](9.2,3.3){$\beta^2$}
	\uput[-135](9.2,1.7){$\beta^2$}
\end{pspicture}
\end{center}
\caption{The general diagram featuring different realizations of the $\frac{\widehat{\mathfrak{osp}}(N|2)_w}{\widehat{\mathfrak{osp}}(N-1|2)_w}$ coset conformal field theory.}
\label{osp(N|2)_cft_screening_set_both_realizations}
\end{figure}

In the case of $OSp(7|2)$, the pattern repeats in Fig. \ref{osp(7|2)_cft_screening_set_both_realizations}, except that the top features a diagram in the shape of the Dynkin diagram for $\mathfrak{osp}(7|2)$, and the bottom features one for $\mathfrak{osp}(6|2)$
\noindent and again, by applying fermionic reflections, we obtain all three realizations mentioned above in Figs. \ref{osp(7|2)_cft_screening_set_first_realization}, \ref{osp(7|2)_cft_screening_set_second_realization}, and \ref{osp(7|2)_cft_screening_set_third_realization} of the $\frac{\widehat{\mathfrak{osp}}(7|2)_w}{\widehat{\mathfrak{osp}}(6|2)_w}$ coset conformal field theory from the screening sets in the dashed and dotted boxes. Summarizing the above, we can write a general scheme in Fig. \ref{osp(N|2)_cft_screening_set_both_realizations} to obtain the screening set of the $\frac{\widehat{\mathfrak{osp}}(2n+1|2)_w}{\widehat{\mathfrak{osp}}(2n|2)_w}$ coset conformal field theory
\noindent where at the lower level we have $n$ screenings to the left of the fermionic screening, and at the upper level we also have $n$ screenings to the right of the fermionic screening. Our main statement is that the screening systems enclosed in the dashed and dotted boxes define the $\frac{\widehat{\mathfrak{osp}}(2n+1|2)_w}{\widehat{\mathfrak{osp}}(2n|2)_w}$ coset conformal field theory, while the integrable deformations of this theory describe the $\eta$-deformed $OSp(N|2m)$ sigma models. Furthermore, it has been verified that this hypothesis works for all cases studied in detail in \cite{Alfimov:2020jpy} and is fully consistent with the method of obtaining screening sets via the blow-up transformations described therein. 

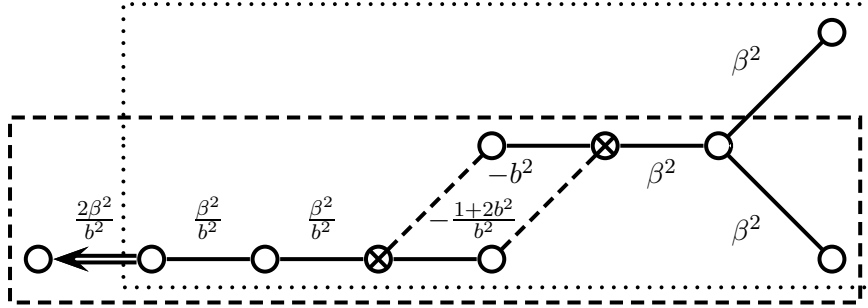
\begin{figure}[H]
\begin{center}
\begin{pspicture}(0.5,0.5)(12,4.5)
	\pscircle[linewidth=1.5pt](1,1){0.2}
    \pscircle[linewidth=1.5pt](2.5,1){0.2}
    \pscircle[linewidth=1.5pt](4,1){0.2}
    \pscircle[linewidth=1.5pt](5.5,1){0.2}
    \psline[linewidth=1.5pt](5.37,0.87)(5.63,1.13)
    \psline[linewidth=1.5pt](5.37,1.13)(5.63,0.87)
    \pscircle[linewidth=1.5pt](7,1){0.2}
    \pscircle[linewidth=1.5pt](7,2.5){0.2}
    \pscircle[linewidth=1.5pt](8.5,2.5){0.2}
    \psline[linewidth=1.5pt](8.37,2.37)(8.63,2.63)
    \psline[linewidth=1.5pt](8.37,2.63)(8.63,2.37)
    \pscircle[linewidth=1.5pt](10,2.5){0.2}
    \pscircle[linewidth=1.5pt](11.5,1){0.2}
    \pscircle[linewidth=1.5pt](11.5,4){0.2}
	\psline[linewidth=1.5pt,doubleline=true,doublesep=1pt]{<-}(1.2,1)(2.3,1)
	\psline[linewidth=1.5pt](2.7,1)(3.8,1)
	\psline[linewidth=1.5pt](4.2,1)(5.3,1)
	\psline[linewidth=1.5pt](5.7,1)(6.8,1)
	\psline[linewidth=1.5pt,linestyle=dashed](5.63,1.13)(6.87,2.37)
	\psline[linewidth=1.5pt,linestyle=dashed](7.13,1.13)(8.37,2.37)
	\psline[linewidth=1.5pt](7.2,2.5)(8.3,2.5)	
	\psline[linewidth=1.5pt](8.7,2.5)(9.8,2.5)
	\psline[linewidth=1.5pt](10.13,2.63)(11.37,3.87)
	\psline[linewidth=1.5pt](10.13,2.37)(11.37,1.13)
	\psframe[linewidth=1.5pt,linestyle=dashed](0.6,0.4)(11.9,2.9)
	\psframe[linewidth=1.5pt,linestyle=dotted](2.1,0.6)(12.1,4.4)
	\uput[90](1.75,1.1){$\frac{2\beta^2}{b^2}$}
	\uput[90](3.25,1.1){$\frac{\beta^2}{b^2}$}
	\uput[90](4.75,1.1){$\frac{\beta^2}{b^2}$}
	\uput[90](6.75,1.1){$-\frac{1+2b^2}{b^2}$}
	\uput[-90](7.25,2.5){$-b^2$}
	\uput[-90](9.25,2.5){$\beta^2$}
	\uput[135](10.7,3.3){$\beta^2$}
	\uput[-135](10.7,1.7){$\beta^2$}
\end{pspicture}
\end{center}
\caption{The general diagram featuring different realizations of the $\frac{\widehat{\mathfrak{osp}}(7|4)_w}{\widehat{\mathfrak{osp}}(6|4)_w}$ coset conformal field theory.}
\label{osp(N|2m)_cft_screening_set_both_realizations}
\end{figure}

Now let us consider the case where the value of $m$ is different from $1$. We have observed that the $\frac{\widehat{\mathfrak{osp}}(7|4)_w}{\widehat{\mathfrak{osp}}(6|4)_w}$ coset conformal field theory and, thus, the $OSp(7|4)$ sigma model can be obtained from the diagrams in Fig. \ref{osp(N|2m)_cft_screening_set_both_realizations},
\noindent enclosed in the dashed and dotted boxes, respectively, taking into account that from the screenings enclosed between two fermionic screenings, we select only one from each level of the diagram (the top and bottom ones are mutually dual to each other).

Based on the information above, we can propose a hypothesis on the structure of the screening charges for arbitrary $N$ and $m$:
\begin{figure}[H]
\begin{center}
\begin{pspicture}(0.5,0.5)(14.5,4.5)
	\pscircle[linewidth=1.5pt](1,1){0.2}
    \pscircle[linewidth=1.5pt](2.5,1){0.2}
    \pscircle[linewidth=1.5pt](4,1){0.2}
    \pscircle[linewidth=1.5pt](5.5,1){0.2}
    \psline[linewidth=1.5pt](5.37,0.87)(5.63,1.13)
    \psline[linewidth=1.5pt](5.37,1.13)(5.63,0.87)
    \pscircle[linewidth=1.5pt](7,1){0.2}
    \pscircle[linewidth=1.5pt](7,2.5){0.2}
    \pscircle[linewidth=1.5pt](8.5,1){0.2}
    \pscircle[linewidth=1.5pt](8.5,2.5){0.2}
    \pscircle[linewidth=1.5pt](10,2.5){0.2}
    \psline[linewidth=1.5pt](9.87,2.37)(10.13,2.63)
    \psline[linewidth=1.5pt](9.87,2.63)(10.13,2.37)
    \pscircle[linewidth=1.5pt](11.5,2.5){0.2}
    \pscircle[linewidth=1.5pt](13,1){0.2}
    \pscircle[linewidth=1.5pt](13,4){0.2}
	\psline[linewidth=1.5pt,doubleline=true,doublesep=1pt]{<-}(1.2,1)(2.3,1)
	\psline[linewidth=1.5pt](2.7,1)(3.0,1)
	\psline[linewidth=1.5pt,linestyle=dotted](3.0,1)(3.5,1)
	\psline[linewidth=1.5pt](3.5,1)(3.8,1)
	\psline[linewidth=1.5pt](4.2,1)(5.3,1)
	\psline[linewidth=1.5pt](5.7,1)(6.8,1)
	\psline[linewidth=1.5pt](7.2,1)(7.5,1)
	\psline[linewidth=1.5pt,linestyle=dotted](7.5,1)(8.0,1)
	\psline[linewidth=1.5pt](8.0,1)(8.3,1)
	\psline[linewidth=1.5pt,linestyle=dashed](5.63,1.13)(6.87,2.37)
	\psline[linewidth=1.5pt,linestyle=dashed](8.63,1.13)(9.87,2.37)
	\psline[linewidth=1.5pt](7.2,2.5)(7.5,2.5)
	\psline[linewidth=1.5pt,linestyle=dotted](7.5,2.5)(8.0,2.5)
	\psline[linewidth=1.5pt](8.0,2.5)(8.3,2.5)
	\psline[linewidth=1.5pt](8.7,2.5)(9.8,2.5)	
	\psline[linewidth=1.5pt](10.2,2.5)(10.5,2.5)
	\psline[linewidth=1.5pt,linestyle=dotted](10.5,2.5)(11.0,2.5)
	\psline[linewidth=1.5pt](11.0,2.5)(11.3,2.5)
	\psline[linewidth=1.5pt](11.63,2.63)(12.87,3.87)
	\psline[linewidth=1.5pt](11.63,2.37)(12.87,1.13)
	\psframe[linewidth=1.5pt,linestyle=dashed](0.6,0.4)(13.4,2.9)
	\psframe[linewidth=1.5pt,linestyle=dotted](2.1,0.6)(13.6,4.4)
	\uput[90](1.75,1.1){$\frac{2\beta^2}{b^2}$}
	\uput[90](4.75,1.1){$\frac{\beta^2}{b^2}$}
	\uput[90](6.75,1.1){$-\frac{1+2b^2}{b^2}$}
	\uput[-90](8.85,2.5){$-b^2$}
	\uput[-45](12.3,3.5){$\beta^2$}
	\uput[45](12.3,1.6){$\beta^2$}
\end{pspicture}
\end{center}
\caption{The general diagram featuring different realizations of the $\frac{\widehat{\mathfrak{osp}}(N|2m)_w}{\widehat{\mathfrak{osp}}(N-1|2m)_w}$ coset conformal field theory.}
\end{figure}
The general picture can be presented as follows: we draw the Dynkin diagrams corresponding to the Lie superalgebras $\mathfrak{osp}(2n+1|2m)$ and $\mathfrak{osp}(2n|2m)$ one above the other in such a way that their parts of length $m-1$ are aligned vertically, with the top and bottom screenings being dual to each other. Then, the parts of this diagram enclosed in the dashed and dotted boxes define the screening systems for the $\frac{\widehat{\mathfrak{osp}}(2n+1|2m)_w}{\widehat{\mathfrak{osp}}(2n|2m)_w}$ coset conformal field theory, which, as we conjecture, can be reduced to a purely fermionic screening system in the general case by applying fermionic reflections. Note that in the middle part of the diagram of length $m-1$, we choose only one screening from each dual pair. The investigation of this screening system in the general case will be the subject of our further study.

%
%
%
%
%
%
%
%
%
%
%
%
%
%


\section{Conclusion}

In conclusion, let us note that we have succeeded in obtaining the following results. First, a new perspective on the method for systematically generating a set of screening charges for deformed $O(N)$ sigma models is presented. Second, the screening charge system that defines the integrable structure of the $OSp(N|2)$ sigma model has been constructed. Using it, we demonstrated how to reconstruct the sigma model action in the deep ultraviolet regime for the cases of $OSp(5|2)$ and $OSp(7|2)$. Furthermore, using our screening charge system, we wrote down the dual Toda-type model for the $OSp(5|2)$ and $OSp(7|2)$ cases. Third, we proposed a conjecture on the method for constructing screening charge sets for arbitrary deformed $OSp(N|2m)$ sigma models.

Based on the obtained results, we can formulate several promising tasks for the future. Namely, it would be interesting to find a screening charge system for a wider class of integrable sigma models. For instance, one could attempt to adapt the dual description for sigma models with a non-compact field space \cite{Basso:2018agi}, as well as to incorporate reparameterization invariance into the dual description.

\section{Acknowledgments}

The author is grateful to Alexey Litvinov, Mikhail Pavlov, and especially Boris Feigin for fruitful and exciting discussions on the topic of the paper. The author's work was supported by the Russian Science Foundation grant No. 25-72-10177.

\bibliographystyle{MyStyle}
\bibliography{MyBib}

\providecommand{\href}[2]{#2}\begingroup\raggedright\begin{thebibliography}{10}

\bibitem{Fateev:1992tk}
V.~A. Fateev, E.~Onofri, and A.~B. Zamolodchikov, {\it {The Sausage model
  (integrable deformations of O(3) sigma model)}},  {\em Nucl. Phys.} {\bf
  B406} (1993) 521--565.

\bibitem{Fateev:1996ea}
V.~A. Fateev, {\it {The sigma model (dual) representation for a two-parameter
  family of integrable quantum field theories}},  {\em Nucl. Phys.} {\bf B473}
  (1996) 509--538.

\bibitem{Litvinov:2018bou}
A.~V. Litvinov and L.~A. Spodyneiko, {\it {On dual description of the deformed
  $O(N)$ sigma model}},  {\em JHEP} {\bf 11} (2018) 139,
  [\href{http://xxx.lanl.gov/abs/1804.0708}{{\tt arXiv:1804.0708}}].

\bibitem{Bazhanov:1984gu}
V.~V. Bazhanov, {\it {Trigonometric Solution of Triangle Equations and
  Classical Lie Algebras}},  {\em Phys. Lett. B} {\bf 159} (1985) 321--324.

\bibitem{Jimbo:1985ua}
M.~Jimbo, {\it {Quantum r Matrix for the Generalized Toda System}},  {\em
  Commun. Math. Phys.} {\bf 102} (1986) 537--547.

\bibitem{Wakimoto:1986gf}
M.~Wakimoto, {\it {Fock representations of the affine lie algebra A1(1)}},
  {\em Commun. Math. Phys.} {\bf 104} (1986) 605--609.

\bibitem{Litvinov:2016mgi}
A.~Litvinov and L.~Spodyneiko, {\it {On W algebras commuting with a set of
  screenings}},  {\em JHEP} {\bf 11} (2016) 138,
  [\href{http://xxx.lanl.gov/abs/1609.0627}{{\tt arXiv:1609.0627}}].

\bibitem{Hoare:2011wr}
B.~Hoare, T.~J. Hollowood, and J.~L. Miramontes, {\it {q-Deformation of the
  $AdS_5 x S^5$ Superstring S-matrix and its Relativistic Limit}},  {\em JHEP}
  {\bf 03} (2012) 015, [\href{http://xxx.lanl.gov/abs/1112.4485}{{\tt
  arXiv:1112.4485}}].

\bibitem{Alfimov:2020jpy}
M.~Alfimov, B.~Feigin, B.~Hoare, and A.~Litvinov, {\it {Dual description of
  $\eta$-deformed OSP sigma models}},  {\em JHEP} {\bf 12} (2020) 040,
  [\href{http://xxx.lanl.gov/abs/2010.1192}{{\tt arXiv:2010.1192}}].

\bibitem{Arutyunov:2013ega}
G.~Arutyunov, R.~Borsato, and S.~Frolov, {\it {S-matrix for strings on
  $\eta$-deformed AdS5 x S5}},  {\em JHEP} {\bf 04} (2014) 002,
  [\href{http://xxx.lanl.gov/abs/1312.3542}{{\tt arXiv:1312.3542}}].

\bibitem{Arutyunov:2015qva}
G.~Arutyunov, R.~Borsato, and S.~Frolov, {\it {Puzzles of $\eta$-deformed
  AdS$_5 \times$ S$^5$}},  {\em JHEP} {\bf 12} (2015) 049,
  [\href{http://xxx.lanl.gov/abs/1507.0423}{{\tt arXiv:1507.0423}}].

\bibitem{Arutyunov:2015mqj}
G.~Arutyunov, S.~Frolov, B.~Hoare, R.~Roiban, and A.~A. Tseytlin, {\it {Scale
  invariance of the $\eta$-deformed $AdS_5\times S^5$ superstring, T-duality
  and modified type II equations}},  {\em Nucl. Phys. B} {\bf 903} (2016)
  262--303, [\href{http://xxx.lanl.gov/abs/1511.0579}{{\tt arXiv:1511.0579}}].

\bibitem{Friedan:1980jf}
D.~Friedan, {\it {Nonlinear Models in Two Epsilon Dimensions}},  {\em Phys.
  Rev. Lett.} {\bf 45} (1980) 1057.

\bibitem{Friedan:1980jm}
D.~H. Friedan, {\it {Nonlinear Models in Two + Epsilon Dimensions}},  {\em
  Annals Phys.} {\bf 163} (1985) 318.

\bibitem{Coleman:1974bu}
S.~R. Coleman, {\it {The Quantum Sine-Gordon Equation as the Massive Thirring
  Model}},  {\em Phys. Rev. D} {\bf 11} (1975) 2088.

\bibitem{Mandelstam:1975hb}
S.~Mandelstam, {\it {Soliton Operators for the Quantized Sine-Gordon
  Equation}},  {\em Phys. Rev. D} {\bf 11} (1975) 3026.

\bibitem{Baseilhac:1998eq}
P.~Baseilhac and V.~A. Fateev, {\it Expectation values of local fields for a
  two-parameter family of integrable models and related perturbed conformal
  field theories},  {\em Nucl. Phys.} {\bf B532} (1998) 567--587,
  [\href{http://xxx.lanl.gov/abs/hep-th/9906010}{{\tt hep-th/9906010}}].

\bibitem{Alfimov:2021sir}
M.~Alfimov and A.~Litvinov, {\it {On loop corrections to integrable 2D sigma
  model backgrounds}},  {\em JHEP} {\bf 01} (2022) 043,
  [\href{http://xxx.lanl.gov/abs/2110.0541}{{\tt arXiv:2110.0541}}].

\bibitem{Fateev:2018yos}
V.~A. Fateev and A.~V. Litvinov, {\it {Integrability, Duality and Sigma
  Models}},  {\em JHEP} {\bf 11} (2018) 204,
  [\href{http://xxx.lanl.gov/abs/1804.0339}{{\tt arXiv:1804.0339}}].

\bibitem{Basso:2018agi}
B.~Basso and D.-l. Zhong, {\it {Continuum limit of fishnet graphs and AdS sigma
  model}},  {\em JHEP} {\bf 01} (2019) 002,
  [\href{http://xxx.lanl.gov/abs/1806.0410}{{\tt arXiv:1806.0410}}].

\end{thebibliography}\endgroup

\end{document}